\documentclass[a4paper,fleqn]{cas-sc}

\usepackage[numbers,sort&compress]{natbib}

\usepackage{algorithm}
\usepackage{algorithmicx}
\usepackage{rotating}
\usepackage[acronym]{glossaries}
\usepackage{booktabs} 
\usepackage{arydshln} 
\usepackage{acronym}
\usepackage{multirow}
\usepackage{tabularx}
\usepackage{pifont} 
\usepackage{float}
\usepackage{xurl}
\usepackage{hyperref}
\usepackage{todonotes}
\usepackage{comment}
\usepackage{xstring} 
\usepackage{cuted} 
\usepackage[section]{placeins}
\usepackage{float}
\usepackage{capt-of} 
\usepackage{circledsteps} 
\usepackage{flafter}

\newcommand{\xmark}{\ding{55}}

\makeatletter
\providecommand{\doi}[1]{%
  \begingroup
    \def\doival{#1}%
    \IfBeginWith{\doival}{doi:}{\StrBehind{\doival}{doi:}[\doival]}{}%
    \let\bibinfo\@secondoftwo
    \urlstyle{rm}%
    \href{https://doi.org/\doival}{\doival}%
  \endgroup
}
\makeatother

\makeatletter
\AtBeginDocument{%
  \expandafter\renewcommand\expandafter\subsection\expandafter{%
    \expandafter\@fb@secFB\subsection
  }%
}
\makeatother

\usepackage[most]{tcolorbox}
\usepackage{xcolor}
\tcbset{
  promptstyle/.style={
    colback=gray!5,
    colframe=gray!50,
    boxrule=0.4pt,
    arc=2pt,
    left=6pt, right=6pt, top=4pt, bottom=4pt,
    fontupper=\small\ttfamily,
    breakable,
    parskip=4pt
  }
}

\newacronym{lm}{LM}{Language Model}
\newacronym{mlm}{MLM}{Masked Language Model}
\newacronym{slm}{SLM}{Small Language Model}
\newacronym{llm}{LLM}{Large Language Model}

\newacronym{col}{COL}{Chain-of-Layer}
\newacronym{hf}{HF}{Hierarchical Format Taxonomy Induction Instruction}

\newcommand{\val}[2]{$#1_{\pm #2}$}
\newcommand{\bval}[2]{$\mathbf{#1}_{\pm \mathbf{#2}}$}

\AtBeginDocument{%
  }

\begin{document}
\let\WriteBookmarks\relax

\shorttitle{Controlled Knowledge Engineering for Taxonomy Induction}
\shortauthors{Limniatis et~al.}

\title[mode=title]{Controlled Knowledge Engineering for Taxonomy Induction with Tool-Grounded LLM Agents and Runtime-Adaptive Policies}

\author[1]{Jonas Limniatis}
\ead{jonas.limniatis@lht.dlh.de}


\affiliation[1]{organization={Lufthansa Technik},
                city={Hamburg},
               citysep={}, 
                postcode={22335}, 
                country={Germany}}

\author[1]{Bennet Schaarmann}
\ead{bennet.schaarmann@lht.dlh.de}



\author[2]{Georg Reuber}
\ead{georg.reuber@lhind.dlh.de}

\affiliation[2]{organization={Lufthansa Industry Solutions},
                city={Norderstedt},
               citysep={},                postcode={22848}, 
                country={Germany}}

\author[3,4,5]{Robert Stahlbock}[orcid=0000-0002-0393-4904]
\cormark[1]
\ead{robert.stahlbock@uni-hamburg.de}

\affiliation[3]{organization={University of Hamburg Business School},
                city={Hamburg},
               citysep={},                postcode={20145}, 
                country={Germany}}                
\affiliation[4]{organization={Nordakademie, University of Applied Sciences},
                city={Elmshorn},
               citysep={},                postcode={25337}, 
                country={Germany}} 
\affiliation[5]{organization={FOM University of Applied Sciences},
                city={Essen},
               citysep={},                postcode={45141}, 
                country={Germany}} 
\cortext[cor1]{Corresponding author}

\begin{abstract}
Taxonomies are structured knowledge representations that support semantic search, recommendation, product catalog organization, and retrieval-augmented generation. Their automated construction from a fixed set of candidate entities requires more than plausible text generation: the resulting hierarchy must be complete, structurally valid, and semantically coherent. Large Language Models (LLMs) are promising for this task, but iterative free-form generation remains vulnerable to omitted or hallucinated entities, invalid hierarchical relations, and inconsistent organizational principles. We introduce TIGA (Taxonomy Induction with Tool-Grounded Agents)\footnote{Code is available at \url{https://github.com/bschaarmann/TIGA}}, a knowledge-engineering architecture that formulates taxonomy induction as an iterative decision process over an explicit state and constrained action space. TIGA combines executable taxonomy-editing tools, an memory that records tool calls, rationales, and outcomes, and a runtime-adaptive organizational policy that the agent revises during inference. An optional global reviewer evaluates the constructed taxonomy against the candidate set and the active policy. We evaluate TIGA on four benchmark taxonomies using two LLM backbones and compare it with single-shot and iterative prompting-based baselines. TIGA achieves the strongest structural-quality results in most evaluated configurations, particularly for recall and F1, but does not dominate on every precision measure and exhibits higher run-to-run variability in some settings. The gains are accompanied by substantially higher token consumption, API-call volume, and inference time. Ablation results show consistent benefits from the policy and interaction-memory components, whereas the effect of global review depends on the dataset and the available revision budget. These findings characterize tool grounding and runtime policy adaptation as promising but costly mechanisms for improving the controlled construction of structured knowledge representations.
\end{abstract}


\maketitle

\section{Introduction}
\label{sec:intro}

Taxonomies, which organize concepts into hierarchical relations, are a foundational knowledge representation for organizing and accessing knowledge at scale. They play a pivotal role in semantic search~\citep{kang-etal-2024-taxonomy, kargupta-etal-2025-taxoadapt}, personalized recommendation~\citep{10.1145/2556195.2556236, huang-2019-taxonomyaware, 9835344, liang-etal-2025-taxonomy}, product catalog categorization~\citep{cheng-etal-2024-e}, and retrieval-augmented generation, where taxonomies support information retrieval and verification of generated outputs~\citep{10.1145/3711896.3736557}. In knowledge-based systems, they provide a central structure for representing domain knowledge, and their structural quality directly affects the reliability of downstream applications. Nevertheless, constructing high-quality taxonomies remains expensive: manual curation is highly labor-intensive, difficult to sustain over time, and inherently limited in scalability. These limitations motivate research on fully automated taxonomy induction~\citep{https://doi.org/10.1111/exsy.13098}. Moreover, even carefully engineered taxonomies can be sensitive to particular choices of dimensions and categorical frameworks used to organize them.

In this work, we focus on entity-centred taxonomy induction: given a root concept and a flat set of candidate entities, the system must construct a complete hierarchy over these entities. This setting differs from corpus-driven taxonomy extraction, where concepts and relations must first be identified from documents, and from taxonomy expansion, where new entities are inserted into an existing taxonomy. In the entity-centred setting considered here, the resulting taxonomy is an organized knowledge representation whose entity scope, semantic relationships, and global structural validity must be preserved throughout the construction process.

Large language models (LLMs) are promising for this task because they can combine lexical knowledge, contextual interpretation, and flexible reasoning when deciding how candidate entities should be organized. However, their native output interface is unconstrained text generation. This creates a mismatch between the capabilities of the model and the requirements of taxonomy construction: a plausible textual response is not necessarily a complete, valid, or globally coherent knowledge representation. Existing approaches to taxonomy induction and related background are reviewed in Section~\ref{sec:related-work}.

In particular, free-form generation remains prone to \emph{hallucinations} and \emph{instruction-following failures}~\citep{10.1145/3627673.3679608, 10.1145/3637528.3671647}. In taxonomy induction, such failures directly affect the resulting knowledge representation: candidate entities may be omitted, unsupported entities may be introduced, and generated parent--child relations may be inconsistent. These errors can lead to incomplete or contaminated taxonomies, semantically inappropriate assignments, and violations of structural properties such as single-parent assignment, acyclicity, connectedness, or root reachability. Consequently, evaluating only the fluency or local plausibility of generated text is insufficient; the induction process must also be assessed by whether it produces a complete and structurally valid representation of the specified candidate space.

A second limitation concerns the representation of the evolving construction process. Although iterative prompting can expose the model to intermediate results, the taxonomy are typically represented as generated text rather than as an explicitly validated system state, while reasoning and decision taken remain implicit. Corrections and organizational principles therefore remain dependent on what the model reproduces in subsequent prompts. This makes it difficult to distinguish accepted state changes from rejected proposals, to preserve domain-specific guidance across iterations, and to enforce structural constraints during construction. The resulting lack of explicit process state can undermine consistency, auditability, and reproducibility in knowledge-based systems.

LLM-based agents offer a suitable architectural basis for addressing this mismatch. Their decision process can be decomposed into repeated perception--decision--action cycles, while tool calling allows individual modifications to be represented as structured operations rather than unrestricted text~\citep{Xi2025, Wang2024, NEURIPS2023_d842425e}. Externalized memory can preserve prior actions and execution feedback, and a revisable policy can make emerging organizational principles explicit across construction steps. These properties are particularly relevant when the target is not a textual answer but a structured knowledge artifact whose intermediate states should remain inspectable. However, agentic systems consume additional tokens and inference calls because of multi-turn tool interaction and context accumulation~\citep{jiang2025adaptation}. Their potential quality benefits must therefore be evaluated against their computational overhead.

These capabilities align naturally with entity-centred taxonomy induction, where constructing a hierarchy from a flat entity set is itself a multi-step, decision-intensive knowledge-engineering process. The task is not merely to generate hierarchical text, but to construct and revise a structured knowledge artifact through a sequence of controlled operations. Recent neighboring work applies LLM-based agent systems to taxonomy and structured-knowledge construction~\citep{tang2026chisagent, lu2026atlas, li2025climb, zeng2025codetaxo}. However, existing approaches do not combine, for entity-centred taxonomy induction, an explicit persistent taxonomy state, an executable action space for structural modifications, externalized memory containing actions, decision rationales and outcomes, and a policy that can be revised during construction.

This leaves open whether such a tool-grounded interaction loop improves the quality and reliability of the resulting knowledge representation, and whether improvements justify its additional computational cost. To address this gap, we introduce TIGA, a novel LLM-based agent framework specifically tailored to entity-centred taxonomy induction. TIGA treats taxonomy construction as a controlled knowledge-engineering process over explicit state rather than as free-form text generation. It maintains an explicit internal state $S_i = (T_i, V_i, \mathcal{P}_i, M_i)$, consisting of the partially constructed taxonomy $T_i$, the set of pending candidate entities $V_i$, a dynamic policy $\mathcal{P}_i$, and an memory $M_i$. Over this state, the agent repeatedly executes iterative perception-decision-action cycles. The action space is realized as invocable tool functions that abstract canonical operations carried out by human experts during taxonomy construction. The memory $M_i$ records recent tool invocations, including the agent’s underlying rationale and the associated execution feedback. By retaining action--rationale--feedback traces, the agent promotes cross-step decision consistency and enables systematic error attribution and targeted revision throughout induction. The dynamic policy $\mathcal{P}_i$ is a tool-editable rule set that makes organizational constraints explicit and revisable during construction.

The study is guided by the following research questions:

\begin{itemize}
\item \textbf{RQ1:} Does TIGA improve the structural quality of induced taxonomies relative to established LLM-based taxonomy-induction baselines across domains, taxonomy sizes, and model backbones?
\item \textbf{RQ2:} How do the dynamic policy, tool-memory, and global reviewer contribute to the taxonomy quality?
\item \textbf{RQ3:} What computational overhead does TIGA incur relative to the baselines, and when is the resulting quality--cost trade-off favorable in practical knowledge-engineering settings?
\end{itemize}

Our main contributions are outlined as follows:
\begin{itemize}
\item We present TIGA, a tool-grounded, LLM-based agent framework for automated taxonomy induction that formulates construction as an iterative decision-making process over an explicit state and constrained action space, thereby enabling more transparent and controllable taxonomy construction in knowledge-based systems.
\item We propose two mechanisms for adaptive taxonomy induction: (i) a dynamic policy that allows the agent to iteratively refine domain-specific organizational principles during inference, and (ii) a tool-memory that records actions, rationales, and outcomes to promote decision consistency and facilitate error diagnosis and correction.
\item We conduct an empirical evaluation of TIGA across four benchmark datasets, multiple taxonomy domains and scales, and two model backbones, using analyses of structural quality and computational cost.
\item We conduct an ablation study to systematically quantify the individual contributions and limitations of the dynamic policy, memory, and reviewer mechanisms.
\item We examine the trade-offs among inference latency, token consumption, API request volume, and taxonomy quality, with the aim of delineating the conditions under which the additional costs are justified.
\end{itemize}

The remainder of the paper is organized as follows. Section~\ref{sec:related-work} reviews prior work on taxonomy induction and LLM-based agents for structured knowledge. Section~\ref{sec:problemdefinition} formalizes the task, Section~\ref{sec:methodology} presents TIGA, Section~\ref{sec:experimen-setup} describes the experimental setup, and Section~\ref{sec:results} reports and discusses the results. The paper ends with the conclusion in Section~\ref{sec:conclusion}.

\section{Related Work}
\label{sec:related-work}

This section positions TIGA with respect to prior work on taxonomy induction and LLM-based agent systems for structured-knowledge construction. We differentiate  between methods that infer taxonomies from textual or structured evidence and methods that construct taxonomies from a predefined candidate entity set. This distinction is critical, as the latter corresponds to the task setting investigated by TIGA and adopted in the experimental evaluation.

\subsection{Taxonomy Induction}
Existing research on taxonomy induction spans manual curation and automated methods. Within automated taxonomy induction, prior work can be broadly grouped into three paradigms: pattern- and structure-based methods, pretrained language-model probing methods, and LLM-based generative methods. These paradigms differ in the information used to infer relations, the degree of supervision required, and the extent to which taxonomy construction is represented as an explicit structural process.

\subsubsection{Pattern- and Structure-Based Methods}
\label{subsubsec:related-pattern-structure}
\citet{hearst1992automatic} introduce lexico-syntactic patterns for extracting hypernym--hyponym relations from unstructured text corpora. Subsequent work moves from hand-crafted patterns to statistical or learned structures, including clustering methods that organize entities directly from text~\citep{grootendorst2022bertopic, 10.5555/3600270.3602558}, and metadata driven taxonomy construction from scratch~\citep{10.1145/3366423.3380259}. \citet{Shen_etal_2017} uses relation extraction to expand an existing taxonomy in depth.

Notable structure-based approaches include belief propagation for integrating relational information between siblings~\citep{bansal-etal-2014-structured} and reinforcement learning for sequential hypernym identification and placement~\citep{mao-etal-2018-end}. Graph-based models such as Graph2Taxo~\citep{shang-etal-2020-taxonomy} further cast induction as cross-domain graph denoising to support domain transfer. Despite their strengths, these methods typically rely on curated lexical resources, document evidence, or substantial domain-specific training data, which limits applicability when only a flat candidate entity set is available.

\subsubsection{Pretrained Language-Model Probing}
\label{subsubsec:related-plm-probing}
A subsequent line of work explores minimally supervised taxonomy induction via pretrained Small Langugage Models (SLMs). \citet{sun-etal-2022-minimally} demonstrates that SLMs encode hypernym–hyponym relations more reliably than earlier baselines, while \citet{chen-etal-2021-constructing, jain-espinosa-anke-2022-distilling} propose MLM-based scoring mechanisms for denoising pairwise hypernym–hyponym relations. These approaches exploit information encoded in pretrained language models while retaining a comparatively explicit relation-scoring formulation. However, they generally focus on identifying or ranking pairwise relations; the subsequent organization of all candidate entities into a complete, globally valid taxonomy remains a separate structural problem.

\subsubsection{LLM-based Methods}
\label{subsubsec:related-llm-methods}
More recently, LLMs have been applied to taxonomy induction via prompting, fine-tuning, and agentic workflows. Their extensive domain knowledge and in-context learning capabilities allow them to infer semantic relationships with minimal or no task-specific parameter optimization. \citet{prompting-vs-ft} categorize existing approaches into two overarching paradigms: \textit{fine-tuning} and \textit{prompt engineering}. \citet{moskvoretskii-etal-2024-large, Moskvoretskii2026} demonstrate that LoRA-based fine-tuning yields improved performance on hypernym--hyponym identification, whereas \citet{prompting-vs-ft} find that prompting can achieve performance comparable to, and in some settings surpassing, that obtained via fine-tuning.

Two prompting-based systems are particularly relevant to our work. TnT-LLM~\citep{10.1145/3637528.3671647} frames taxonomy generation as a zero-shot prompt chain over LLM-generated document summaries, analogous to stochastic optimization of a mixture model: an initial taxonomy is generated from a first mini-batch, iteratively refined by an update prompt over subsequent mini-batches, and finalized by a review prompt.\citet{10.1145/3627673.3679608} propose two complementary taxonomy-induction methods: Hierarchical Format Taxonomy Induction Instruction (HF) generates the entire taxonomy in a single pass, encoding structure through a fixed hierarchical numbering format that makes sibling and ancestor relations explicit. Chain-of-Layer (CoL) builds on HF by decomposing induction into top-down, layer-by-layer entity selection. At each iteration, an MLM-based ensemble ranking filter discards generated parent--child edges that fall outside the child’s top-ranked parent candidates, which limits hallucinated relations and the propagation of errors across layers. These methods provide strong prompting-based reference points for the entity-centred evaluation setting considered in this paper.

Despite recent advances, existing LLM-based approaches generally realize reasoning, decision-making, and structural modification through generated text that is subsequently parsed or filtered. This design creates three interrelated limitations: (i) vulnerability to hallucinated entities and relations \citep{prompting-vs-ft,10.1145/3627673.3679608}; (ii) the reasoning behind each structural decision remains implicit and is not persisted as verifiable state, so decisions cannot be checked or reused across steps, which can yield inconsistent structures during multi-step construction \citep{10.1145/3627673.3679608}; and (iii) reliance on static prompt-defined guidance that does not explicitly persist and revise organizational principles as the taxonomy evolves \citep{10.1145/3637528.3671647}. These limitations are particularly relevant when the output must be a complete and structurally valid knowledge representation rather than a plausible hierarchical text.

The methods discussed above demonstrate that LLMs can induce useful taxonomic relations and structures through prompting or fine-tuning. However, their primary unit of interaction remains the generated response. The following subsection therefore considers a related but broader line of work: LLM-based agent systems for taxonomy and structured-knowledge construction, with particular attention to whether tools, memory, reflection, and persistent state address the limitations identified above.

\subsection{LLM-based Agent Systems for Structured Knowledge}
\label{subsec:related-agentic-work}

Recent neighboring work applies LLM-based agent systems to taxonomy and structured-knowledge construction. These systems demonstrate the value of decomposition, reflection, retrieval, or collaboration, but they differ from TIGA in task definition and in how structural modifications are represented and executed.

CHisAgent~\citep{tang2026chisagent} constructs event taxonomies with role-specialized bottom-up and top-down agents that extract, merge, and insert nodes through free-form generation. It therefore addresses corpus-based event-taxonomy construction and does not provide an executable tool interface over a persistent, tool-mediated taxonomy state. ATLAS~\citep{lu2026atlas} couples prompted Designer and Classifier LLM in a self-corrective loop to derive taxonomies over software repositories. Its task is dimension-splitting taxonomy construction and differs from hypernym induction from a flat candidate entity set.

CLIMB~\citep{li2025climb} builds occupation taxonomies bottom-up from unstructured job postings using a multi-agent reflection loop consisting of a generator and evaluator and using semantic clustering. Although it improves coherence and regional adaptability over prior bottom-up baselines, its critique function is fixed and hand-specified, and taxonomy construction is performed through free-form generation rather than executable tool calls over a persistent state.

CodeTaxo~\citep{zeng2025codetaxo} is closest to our approach in representing taxonomy edits as code-like operations. However, these operations are generated as output text rather than executed through a tool interface. Moreover, CodeTaxo focuses on taxonomy expansion by inserting individual query entities into an existing taxonomy, whereas TIGA constructs a complete taxonomy over a fixed candidate entity set.

Agent-OM~\citep{jiang2024agentom} uses two LLM-based agents with retrieval and matching tools for ontology alignment. It illustrates the use of tool-grounded agents in a related structured-knowledge task, but ontology alignment is distinct from taxonomy induction because the objective is to match entities or concepts across existing ontologies rather than to infer a new hierarchy over a flat candidate set. Soularidis et al.~\citep{Soularidis_Doumanas_Kotis_Vouros_2025} emulate multiple stakeholder roles within a single LLM augmented by retrieval and ReAct-style guidelines. Their system supports collaborative ontology engineering, but the agents do not execute taxonomy operations through a constrained structural action space; the ontology is generated in free-form text.

Taken together, the reviewed studies establish the relevance of LLM-based agents for iterative and structured-knowledge tasks. They also show that taxonomy induction, taxonomy expansion, ontology alignment, and collaborative ontology engineering involve different task definitions and structural requirements. For the entity-centred induction setting considered here, the combination of an explicit persistent taxonomy state, executable structural actions, externalized memory, and a runtime-revisable policy remains insufficiently explored. The following Section formalizes this task setting and its structural requirements.

\section{Task Definition}
\label{sec:problemdefinition}

We formalize taxonomy induction as the construction of a rooted, directed tree that represents hypernym--hyponym relations among a fixed set of candidate entities. Let
$\mathcal{T} = (\mathcal{V}, \mathcal{E})$ denote a taxonomy with root $v_0 \in \mathcal{V}$, where $\mathcal{V}$ is a set of entities and $\mathcal{E} \subseteq \mathcal{V} \times \mathcal{V}$ is a set of directed hypernym--hyponym edges. An edge $(u, v) \in \mathcal{E}$ indicates that $u$ is a direct hypernym of $v$. Accordingly, every non-root entity has exactly one parent, the root has no parent, and the resulting structure is connected and acyclic.

Given a root entity $v_0$ and an initial, flat candidate entity set $V_0 = \{v_1, \dots, v_n\}$, taxonomy induction seeks to construct a taxonomy $\mathcal{T}^* = (\mathcal{V}^*, \mathcal{E}^*)$ such that
\begin{equation}
    \mathcal{V}^* = V_0 \cup \{v_0\},
\end{equation}
and $\mathcal{E}^*$ defines a valid rooted tree over $\mathcal{V}^*$. The root and candidate entities are provided as input, whereas the parent--child relations are not specified and must be inferred by the induction system. The objective is to maximize structural agreement with a gold-standard taxonomy $\mathcal{T}^{\text{gold}}$ defined over $V_0 \cup \{v_0\}$, as quantified by the edge-level and ancestor-level metrics introduced in Section~\ref{sec:experimen-setup}.

Because the candidate set is fixed, the induced structure must contain each entity of $\mathcal{V}^*$ as exactly one node and must not introduce entities outside $\mathcal{V}^*$. Induction is therefore not an open-ended concept-discovery task, but the controlled organization of a given entity set into a valid hierarchical knowledge representation. This distinguishes our setting from corpus-driven taxonomy extraction, where concepts and relations are first identified from documents, and from taxonomy expansion, where new entities are inserted into an existing taxonomy.

Formally, $\mathcal{T}^*$ is a valid rooted tree if (i) $v_0$ has no incoming edge, (ii) every $v \in \mathcal{V}^* \setminus \{v_0\}$ has exactly one parent $u \in \mathcal{V}^*$ with $(u,v) \in \mathcal{E}^*$, and (iii) every entity is reachable from $v_0$. These conditions imply that $\mathcal{T}^*$ is acyclic and that $|\mathcal{E}^*| = |\mathcal{V}^*| - 1$. Validity is defined independently of agreement with the gold taxonomy: an induced structure may be a valid tree while still containing semantically incorrect parent--child relations.

The objective of induction therefore is twofold. First, to construct a complete and structurally valid tree over the specified candidate space. Second, its edges should agree with a gold-standard taxonomy $\mathcal{T}^{\text{gold}}$ over $\mathcal{V}^*$. 

\section{Methodology}
\label{sec:methodology}

In this section, we introduce TIGA, an LLM-based agent framework that formulates entity-centered taxonomy induction as a controlled knowledge-engineering process over an explicit state.
Rather than generating a taxonomy as unrestricted text, TIGA constructs and revises a structured knowledge representation through auditable tool invocations.
The central design principle is to separate the agent's semantic decision-making from the execution of structural modifications: the agent proposes an operation, whereas a tool layer executes it against the current workspace, and returns explicit feedback.

TIGA combines \emph{action-level decomposition}, where each modification is a discrete verifiable tool invocation against the current taxonomy state; an \emph{externalized memory}, which records past actions, rationales, and outcomes to support consistent, error-aware decisions; and a \emph{dynamic policy}, a set of natural-language construction guidelines that the agent revises during execution to adapt to domain characteristics without fine-tuning. Together, these mechanisms treat the evolving taxonomy as an explicit, verifiable knowledge artifact rather than as a sequence of generated text.

Section~\ref{subsec:methodology-context-management} formalizes the state $S_i = (T_i, V_i, \mathcal{P}_i, M_i)$. Section~\ref{subsec:methodology-tools} specifies the action space $\mathcal{A}$ and its transition semantics. Section~\ref{subsec:methodology-agent-execution} details the iterative decision-action procedure and the global LLM-based judge that reviews the constructed taxonomy.

\subsection{Environment State Representation and Agent Context}
\label{subsec:methodology-context-management}

Earlier LLM-based methods frequently struggle with error propagation. This is evident in iterative multi-step pipelines, where structural and semantic errors introduced early persist and condition subsequent decisions ~\citep{10.1145/3637528.3671647}. Although few-shot in-context learning can improve output quality, it does not  these failure modes~\citep{10.1145/3627673.3679608}. In our setting, these include (i)~\emph{instruction loss}, where the LLM deviates from the required hierarchical format and produces free text; (ii)~\emph{relational hallucinations}, where incorrect parent--child assignments are introduced; and (iii)~\emph{entity hallucinations}, where entities outside the candidate set are inserted.

\subsubsection{Internal State} TIGA addresses these failure modes by maintaining an explicit internal state that grounds each action in a verifiable knowledge representation. This design follows the principle that agents must revise their behavior in response to environmental feedback rather than committing to a single upfront plan~\citep{Wang2024}. The decision history $M_i$ externalizes this feedback as memory, enabling future actions to be conditioned on observed outcomes rather than on re-planning from implicit feedback. The policy $\mathcal{P}_i$ extends this loop by making organizational principles explicit and revisable across steps.

Concretely, the state at iteration $i$ is defined as
\begin{equation}
    S_i = (T_i,\; V_i,\; \mathcal{P}_i,\; M_i),
\end{equation}
where $T_i$ denotes the current partial taxonomy and $V_i$ the set of remaining candidate entities. The state evolves through tool invocations whose transition semantics are specified in Section~\ref{subsec:methodology-tools}. We describe the remaining components in the following subsections.

Unlike agent configurations in which salient external dynamics are only partially observable or amenable to control (e.g., web search), the task state in TIGA is entirely specified and modified through the agent’s own tool invocations. Accordingly, state transitions are deterministic given the current state, the invoked action, and the associated tool inputs. This property applies to the execution layer and does not imply that the LLM's action selection is deterministic. It supports reproducibility, auditability, and enables systematic identification and diagnosis of structural errors in the induced taxonomy.

\subsubsection{Dynamic policy}
The policy $\mathcal{P}_i$ is a mutable set of organizational guidelines that the agent maintains and revises during inference. It accumulates organizational principles, naming conventions, structural constraints, and restrictions on tool use. This mechanism enables domain-specific adaptation without requiring fine-tuning or domain-specific prompt engineering. In contrast to static prompt engineering, which imposes behavioral constraints only at initialization, $\mathcal{P}_i$ acts as an \emph{agent-authored control layer} that is iteratively refined via the \texttt{update\_policy} tool (Table~\ref{tab:tools}) in response to the current state. Accordingly, the policy renders domain-specific organizational decisions explicit and persistent throughout the construction process.

\subsubsection{Tool-interactive Memory}
We model the interaction memory as a sliding window containing the $i$ latest tool calls:
\begin{equation}
    M_i = (m_{i-n+1}, m_{i-n+2}, \ldots, m_i),
    \quad
    m_i = (a_i, r_i, o_i),
\end{equation}
where $a_i \in \mathcal{A}$ is the invoked action, $r_i$ the provided agent decision rationale, and $o_i$ the observed outcome---either a successful state update or an explicit error message. The window size $n$ is a hyperparameter. The truncation procedure constrains the contextual window to remain within the model's input capacity while retaining a short-term memory of recent decisions and their associated environmental feedback. Externalizing the trace within $M_i$ supports planning, error-aware self-correction, and iterative policy refinement.

\subsubsection{Agent Context} 
The full information accessible to the agent at iteration $i$ is given by
\begin{equation}
    C_i = (S_i,\; D,\; \mathcal{A}),
\end{equation}
where $D = \{d_1, \ldots, d_5\}$ represents a fixed set of few-shot demonstrations used for in-context learning and $\mathcal{A}$ denotes the available tool set (Section~\ref{subsec:methodology-tools}). Both $D$ and $\mathcal{A}$ stay fixed across iterations; only $S_i$ varies as actions are performed. The demonstrations offer structural priors as reference sub-taxonomies, guiding the agent toward valid hierarchical structures.

\subsection{Constrained Action Space}
\label{subsec:methodology-tools}

Prior iterative multi-step LLM-based taxonomy induction methods elicit LLMs' reasoning capabilities through prompt engineering, chain-of-thought reasoning~\citep{NEURIPS2022_9d560961}, and, in iterative approaches, multi-step review chains~\citep{10.1145/3627673.3679608, 10.1145/3637528.3671647}. The intermediate decision-making process in these approaches, however, remains mostly implicit: individual modifications to the taxonomy structure are neither explicitly specified nor justified, limiting interpretability and structural consistency.

We formulate the \emph{action module} as a space consisting of callable internal tool functions that modify the task environment when invoked~\citep{Wang2024, Xi2025}. Each action is a deterministic, auditable operation on the workspace, conditional on the current state and the supplied tool inputs. The agent is restricted to this predefined tool set rather than generating unrestricted free-form taxonomy content. The tools are well-definedprecisely specified, goal-oriented, and engineered to support the requirements of the taxonomy induction task. To minimize ambiguity, we intentionally keep tool primitives small. Each invocation produces a clearly defined result stored in $M_i$, and operational constraints, encoded in $\mathcal{P}_i$, act as guardrails to prevent invalid modifications. Each tool exposes only the inputs strictly necessary for its use, limiting the agent's options and reducing the likelihood of hallucinated arguments.

The action space is partitioned into two disjoint classes: state-mutating and read-only
\begin{equation}
    \mathcal{A}
    = \underbrace{\{a^{\texttt{add}},\; a^{\texttt{remove}},\;
                    a^{\texttt{policy}}\}}_{\text{state-mutating}}
    \;\cup\;
    \underbrace{\{a^{\texttt{view}}\}}_{\text{read-only}},
\end{equation}
corresponding to the tools described in Table~\ref{tab:tools} for full specifications and input/output schemas. At iteration $i$, the agent selects an admissible action through an LLM-based decision function
\begin{equation}
    a_i = \pi_\theta(C_i), \qquad a_i \in \mathcal{A},
\end{equation}
where agent system $\pi_\theta : \mathcal{C} \to \mathcal{A}$ maps the current context to a single action.

All state-mutating tools must be invoked only when accompanied by an explicit rationale argument that justifies their invocation. By contrast, the read-only tool accepts as input a candidate entity identifier and returns the associated descendant subgraph of $T_i$, while leaving $S_i$ unchanged.

\subsection{Inference via Agent Decision Process}
\label{subsec:methodology-agent-execution}

\begin{figure}[t]
  \centering
  \includegraphics[width=\linewidth]{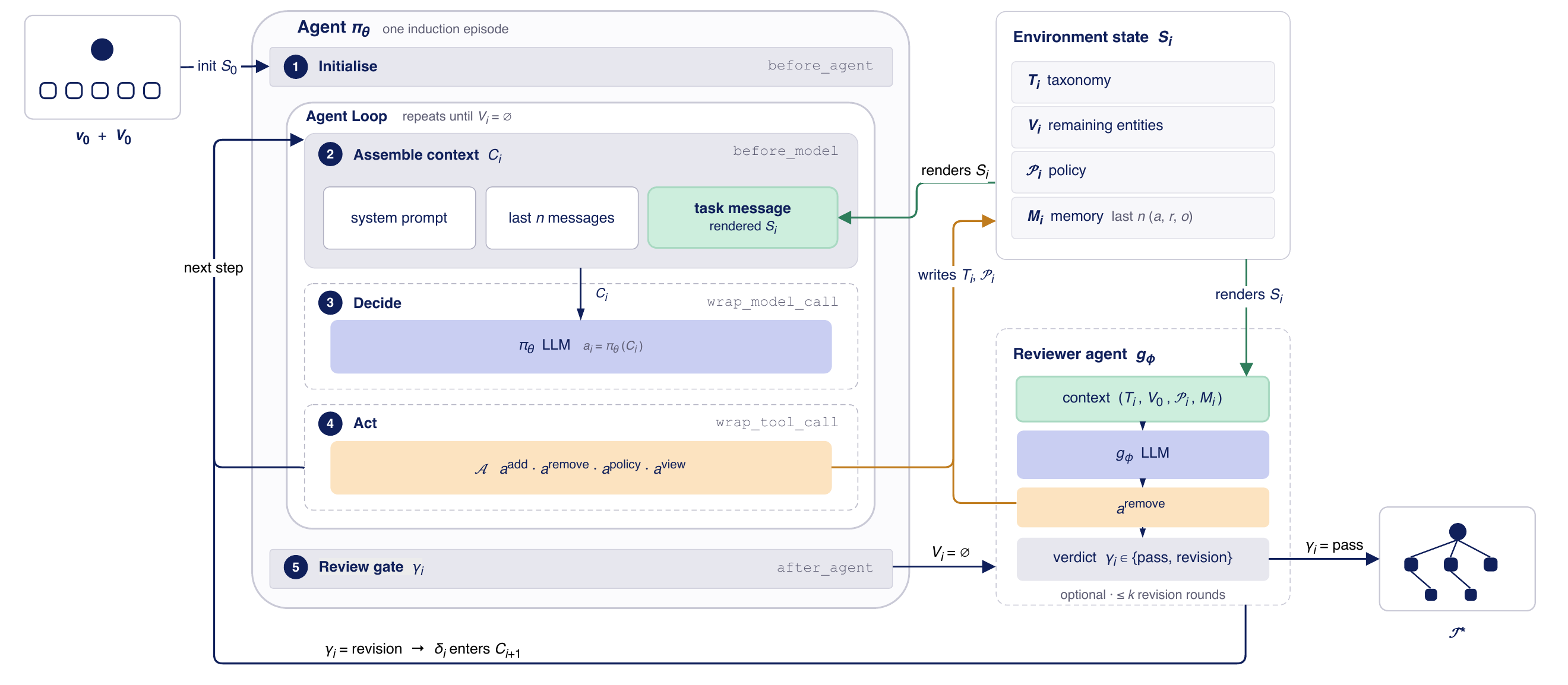}
  \caption{Overview of the TIGA inference workflow. Starting from a root concept $v_0$ and a flat candidate set $V_0$, the agent $\pi_\theta$ executes one induction episode over the environment state
  $S_i=(T_i,V_i,\mathcal{P}_i,M_i)$ (Section~\ref{subsec:methodology-context-management}).
  Each iteration of the agent loop (\Circled{2}--\Circled{4}) assembles the context $C_i$ from a system prompt, the last $n$ messages, and a task message that renders the current state $S_i$; the LLM selects a single tool action $a_i \in \mathcal{A}$ (Section~\ref{subsec:methodology-tools}), whose successful execution writes back to $T_i$ and $\mathcal{P}_i$. The loop repeats until $V_i=\emptyset$ (i.e., every candidate entity has been placed), at which the reviewer agent $g_\phi$ is being invoked over $(T_i,V_0,\mathcal{P}_i,M_i)$ (Section~\ref{subsec:methodology-agent-execution}). A verdict $\gamma_i=\textsc{pass}$ terminates inference and emits the final taxonomy $\mathcal{T}^{*}$, whereas $\gamma_i=\textsc{revision}$ prepends the rationale $\delta_i$ to $C_{i+1}$ and re-enters the loop for at most $k$ rounds. The reviewer is restricted to $a^{\texttt{remove}}$ and therefore only prunes $T_i$. Grey monospace labels denote the corresponding middleware hooks of the
  implementation (Section~\ref{subsubsec:implementation}).}
  \label{fig:framework-overview}
\end{figure}

Fig.~\ref{fig:framework-overview} presents a high-level overview of the framework. Inference unfolds as an iterative loop in which the agent repeatedly selects a tool action, perceives the resulting outcome, and interacts with the environment. This process continues until every candidate entity has been inserted into the taxonomy, after which a prompting-based LLM reviewer decides whether to terminate the process or initiate a revision phase.

\subsubsection{Iterative State Transition}
At iteration $i$, the agent invokes a tool call $(a_i, r_i)$ which is executed against the current environment state, producing an outcome $o_i \in \{\textsc{ok}, \textsc{err}\}$ together with any tool return value. The workspace state then transitions according to
\begin{equation}
S_{i+1} = f(S_i, a_i, r_i, o_i)
        = (T_{i+1}, V_{i+1}, \mathcal{P}_{i+1}, M_{i+1}).
\end{equation}
Successful state-mutating actions update the corresponding component of the state; unsuccessful actions ($o_i = \textsc{err}$) leave $T$, $V$, and $\mathcal{P}$ unchanged. Read-only actions leave all components except $M$ unchanged. In every case, the invocation is appended to the memory. Formally:
\begin{align}
T_{i+1} &=
\begin{cases}
\texttt{add}(T_i, e) & a_i{=}a^{\texttt{add}},\; o_i{=}\textsc{ok} \\
\texttt{remove}(T_i, e) & a_i{=}a^{\texttt{remove}},\; o_i{=}\textsc{ok} \\
T_i & \text{otherwise}
\end{cases}
\\[4pt]
V_{i+1} &=
\begin{cases}
V_i \setminus \{e\} & a_i{=}a^{\texttt{add}},\; o_i{=}\textsc{ok} \\
V_i \cup \{e\} & a_i{=}a^{\texttt{remove}},\; o_i{=}\textsc{ok} \\
V_i & \text{otherwise}
\end{cases}
\\[4pt]
\mathcal{P}_{i+1} &=
\begin{cases}
\mathcal{P}' & a_i{=}a^{\texttt{policy}},\; o_i{=}\textsc{ok} \\
\mathcal{P}_i & \text{otherwise}
\end{cases}
\\[4pt]
M_{i+1} &= \mathrm{tail}_n\big(M_i \,\Vert\, (a_i, r_i, o_i)\big)
\end{align}
In this context, $e$ denotes the target entity provided as an argument to $a^{\texttt{add}}$ or $a^{\texttt{remove}}$, and $\mathcal{P}'$ represents the updated policy provided as an argument to $a^{\texttt{policy}}$. The outcome $o_i = \textsc{err}$ is generated when the preconditions associated with a tool are not satisfied (e.g., when adding an entity under a non-existent parent). In such cases, a descriptive error message is included in $o_i$ to support subsequent inspection and systematic handling by the agent.

\subsubsection{Review}

When all candidate entities have been processed, indicated by $V_i = \emptyset$, the inference loop invokes an LLM-based global critic:
\begin{equation}
(\gamma_i, \delta_i) = g_\phi(T_i, V_0, \mathcal{P}_i, M_i),
\end{equation}
which returns a verdict $\gamma_i \in \{\textsc{pass}, \textsc{revision}\}$ and a natural-language rationale $\delta_i$. The critic evaluates $T_i$ against two criteria: fidelity to the original candidate set $V_0$, ensuring that the taxonomy reflects the intended candidate space, and compliance with the constraints encoded in $\mathcal{P}_i$. The verdict $\gamma_i = \textsc{pass}$ indicates that both criteria are satisfied and terminates the loop. The verdict $\gamma_i = \textsc{revision}$ indicates that revision is required. In the latter case, $\delta_i$ is prepended to the agent's context $C_{i+1}$, and the loop resumes with $V_i$ repopulated by the entities flagged for reconsideration.

Unlike local self-correction within the inference loop, which relies on tool outcomes recorded in $M_i$, the global reviewer operates over the complete final workspace state. It can therefore identify inconsistencies that span multiple local decisions, such as violated depth conventions, semantically incoherent subtrees, or entity-set drift relative to $V_0$.

\section{Experimental Setup}
\label{sec:experimen-setup}
This section describes the experimental design used to evaluate TIGA as a controlled knowledge-engineering system for entity-centred taxonomy induction. The evaluation considers three complementary dimensions: (i) the structural quality of the induced taxonomies, (ii) the contribution of individual architectural components, and (iii) the computational implications of tool-grounded construction. We evaluate TIGA against established LLM-based taxonomy induction baselines using benchmark datasets.

\subsection{Datasets}
To assess how well TIGA generalizes across different domains and taxonomy scales, we consider four widely used benchmark datasets. These benchmarks vary in domain, the number of candidate entities, taxonomy depth, and branching factor, thereby allowing us to evaluate the robustness of the proposed knowledge-construction process under heterogeneous structural settings. Table~\ref{tab:dataset_stats} summarizes their main properties.

\paragraph{WordNet.}
We adopt the WordNet sub-taxonomy collection introduced by \citet{bansal-etal-2014-structured}, which consists of 761 non-overlapping height-3 subtrees extracted from the WordNet noun hierarchy. Each subtree contains 10--50comprises 10–50 entities (e.g., animal taxa and artifact categories). We adopt the standard split into 533 training, 114 development, and 114 test sub-taxonomies and, following \citet{10.1145/3627673.3679608}, report test-set results. Overall, the sub-taxonomies are shallow and compact, with 20.5 entities on average (mean depth 3.0; mean branching factor 3.8). The root and candidate entities are provided to the induction methods, whereas the gold parent--child relations are withheld and used only for evaluation.

\paragraph{DBLP.}
DBLP~\citep{10.1145/3219819.3220115}is a computer science taxonomy constructed from a corpus of approximately $156{,}000$ paper abstracts. Using the evaluation protocol of \citet{10.1145/3627673.3679608}, we assess performance on five sub-taxonomies that contain 90.8 entities on average, with mean depth 3.5 and mean branching factor 12.6. As in the other benchmarks, the induction methods receive the root and candidate entity set but not the gold parent--child relations.

\paragraph{TExEval-2.}
TExEval-2~\citep{task13semeval2016} is the SemEval-2016 Task 13 benchmark for taxonomy extraction, in which hypernym--hyponym relations must be identified among a given list of domain-specific terms and organized into a domain taxonomy. We use the English gold-standard taxonomies covering the Environment, Food, and Science domains. As the task is organized in an unsupervised setting, no training data are provided; accordingly, no benchmark-specific few-shot examples are provided. We evaluate on 18 gold-standard sub-taxonomies with 94.2 entities on average and a mean branching factor of 14.1 children per non-leaf node.

\paragraph{Google Product Taxonomy (GProd).}
The Google Product Taxonomy is a hierarchical product classification scheme used in Google Merchant Center and Google Shopping.\footnote{See \url{https://www.google.com/basepages/producttype/taxonomy-with-ids.en-US.txt}. We use the version published on 2024-12-12.} In contrast to the scholarly benchmarks above, GProd captures practitioner-oriented structures found in real-world product catalogs. The full taxonomy contains approximately 5{,}500 nodes across 21 top-level categories.
Since the complete taxonomy is too large to fit within the context-window capacity of the evaluated methods, we use a standard sub-sampling procedure: we truncate the hierarchy at depth $4$ (with the top-level categories at depth $1$) and sample 10 sub-taxonomies, each rooted at a distinct top-level category. The resulting sub-taxonomies contain between 89 and 126 entities each. Truncation is applied before induction, so categories beyond depth $4$ are excluded from both the candidate sets and the corresponding gold structures.

\begin{table}[!htbp]
\centering
\caption{Statistics of the evaluation taxonomies. Values are
mean\,/\,max across the listed number of taxonomies per dataset.
Branching denotes children per non-leaf node. Datasets marked
$\dagger$ are pruned at depth $n$ from their full versions to fit
context-budget constraints.}
\label{tab:dataset_stats}
\small
\renewcommand{\arraystretch}{1.15}
\setlength{\tabcolsep}{3pt}
\begin{tabular*}{\columnwidth}{@{\extracolsep{\fill}} l c c c c c c @{}}
\toprule
\textbf{Dataset} & \textbf{\#tax} & $|V|$ & $|E|$ & \textbf{Branch} & \textbf{Max d} \\
\midrule
WordNet              & 114 & 20.5\,/\,46    & 19.5\,/\,45    & 3.8\,/\,9.3    & 3.0\,/\,3 \\
TExEval-2$^\dagger$         & 18  & 94.2\,/\,140   & 94.4\,/\,140   & 14.1\,/\,77.0  & 3.5\,/\,5 \\
DBLP$^\dagger$       & 5   & 90.8\,/\,95    & 89.8\,/\,94    & 12.6\,/\,14.3  & 2.8\,/\,3  \\
GProd$^\dagger$ & 10  & 90.2\,/\,127   & 89.2\,/\,126   & 6.8\,/\,10.7   & 2.8\,/\,3 \\
\bottomrule
\end{tabular*}
\end{table}

\subsection{Baselines}
Given that TIGA targets entity-centric taxonomy induction, we restrict our comparison to methods that operate under an equivalent input configuration. All compared methods receive the root and the same candidate entity set for a given instance, while the gold parent--child relations are withheld. We do not include corpus-driven methods, as they address a different problem formulation (Section~\ref{sec:problemdefinition}): such methods infer concepts and relations from document collections, whereas the present study evaluates the organization of a fixed candidate entity set.

\paragraph{Single-shot LLM prompting.} Hierarchical Format Taxonomy Induction Instruction \textbf{(HF)}~\citep{10.1145/3627673.3679608}: prompts the LLM with a hierarchical-format instruction and a set of candidate entities, producing the complete taxonomy in a single generation step.

\paragraph{Iterative LLM prompting.} Chain-of-Layer (\textbf{CoL})~\citep{10.1145/3627673.3679608}: a strong recent prompting-based method that generates the taxonomy top-down, layer by layer, applying an ensemble-based masked-LM ranking filter at each layer to reduce hallucinated relations.

The baselines differ from TIGA in how they represent intermediate decisions and modify the evolving taxonomy. HF produces the complete structure in a single generation step, whereas CoL performs iterative layer-wise generation and filtering. In both cases, taxonomy modifications remain part of the generated-output or post-processing pipeline rather than being executed as tool calls over the persistent workspace used by TIGA.

\subsection{Evaluation Metrics}
\label{sec:metrics}

We evaluate taxonomy induction along two complementary dimensions: structural quality and computational cost. Structural quality is quantified using precision, recall, and F1 under two complementary protocols: \emph{edge-level} scores, which capture agreement on direct parent--child relations, and \emph{ancestor-level} scores, which evaluate the transitive closure of ancestor--descendant pairs and thus reflect global hierarchical consistency beyond individual edges.

Because precision and recall do not fully establish whether an output is a valid knowledge representation, we additionally record candidate coverage, the rate of extraneous entities, and the proportion of outputs satisfying the required rooted-tree constraints. These constraints include single-parent assignment for every non-root entity, root reachability, connectedness, and acyclicity. The resulting validity measures distinguish structural validity from semantic agreement with the gold taxonomy.

To address computational cost, we report wall-clock inference time, the total number of model/API calls, and total token consumption, separated into prompt and completion tokens where available. For TIGA, we additionally record the number of tool invocations, policy updates, reviewer calls, and revision rounds. These measures characterize the computational overhead of fine-grained, tool-mediated knowledge construction. Unless explicitly stated otherwise, we use \emph{computational cost} to refer to these measured proxies and do not equate them with monetary API cost.

\subsection{Implementation Details}
\label{subsubsec:implementation}

For TIGA and all LLM-based baselines, we evaluate two backbones: a proprietary model (\texttt{GPT-5.4-mini}\footnote{API version: \texttt{GPT-5.4-mini-2026-03-17}}) and an open-weight model (\texttt{Gemma4-31b}). Both backbones receive identical task inputs, differing only in the method-specific prompting templates.

All methods are evaluated at temperature $0.0$. Each configuration is executed in five independent trials under identical conditions; we report the mean $\pm$ standard deviation across trials to quantify run-to-run variability. Consequently, the reported standard deviations reflect variability induced by the full inference pipeline, including any nondeterminism attributable to the model provider and/or the execution environment.

TIGA integrates four invocable tools (Table~\ref{tab:tools}).
To bound inference cost and limit the effective context length, the live dialogue is truncated to the two most recent messages ($n=2$). Tool invocations occurring outside this window are serialized into our tool-interaction unbounded memory $M_i$, and subsequently re-injected into the prompt for the next task. Consequently, $n$ constrains only the active conversational transcript rather than the persistently retained interaction memory. This design choice operationally differentiates the transient context window from the persistent interaction log.

To further constrain inference-related computational cost, the review procedure is capped at a maximum of three revision iterations. The reviewer module is deactivated for the baseline results reported in Section~\ref{sec:baseline} and activated for the experiments in Section~\ref{sec:ablation-study}.

HF and CoL were evaluated under zero-shot and 5-shot conditions using the authors’ publicly released code and associated few-shot demonstrations~\citep{10.1145/3627673.3679608}. We further extended the provided implementations to record inference latency and token usage. Zero-shot and 5-shot evaluations are reported as distinct experimental configurations.

\section{Results and Discussion}
\label{sec:results}

\subsection{Taxonomy Structural Quality}
\label{sec:baseline}

Table~\ref{tab:main_results_compact} reports $F1_a$ and $F1_e$ scores across the four benchmarks, the two backbone models, and the evaluated shot settings; the complete precision--recall breakdown under both edge-level and ancestor-level evaluation protocols is provided in Table~\ref{tab:large_results}.

\begin{table}[!htbp]
\centering
\caption{Mean $F1_a$ and $F1_e$ across four benchmarks, both 
backbones, and both shot settings. Values are 
mean$_{\pm\text{std}}$ over 5 random seeds (in percent). Bold 
indicates the best result per column. Full precision/recall 
breakdown appears in Table \ref{tab:large_results}.}
\label{tab:main_results_compact}
\small
\resizebox{\columnwidth}{!}{%
\begin{tabular}{lcccccccc}
\toprule
& \multicolumn{2}{c}{WordNet} & \multicolumn{2}{c}{DBLP} 
& \multicolumn{2}{c}{TExEval-2} & \multicolumn{2}{c}{GProd} \\
\cmidrule(lr){2-3} \cmidrule(lr){4-5} \cmidrule(lr){6-7} \cmidrule(lr){8-9}
\textbf{Model} & $F1_a$ & $F1_e$ & $F1_a$ & $F1_e$ & $F1_a$ & $F1_e$ & $F1_a$ & $F1_e$ \\
\midrule
\multicolumn{9}{l}{\textit{Zero-shot Setting}} \\
\midrule
HF (GPT-5.4-mini) & \val{61.44}{1.76} & \val{50.32}{1.33} & \val{13.28}{1.07} & \val{22.27}{2.36} & \val{28.07}{3.77} & \val{28.31}{3.15} & \val{52.53}{5.68} & \val{38.31}{4.95} \\
HF (Gemma4-31b)   & \val{66.83}{0.03} & \val{55.51}{0.11} & \val{11.49}{0.00} & \val{9.45}{0.00}  & \val{24.38}{0.35} & \val{26.25}{0.28} & \val{48.01}{0.00} & \val{42.80}{0.00} \\
CoL (GPT-5.4-mini)& \val{73.97}{2.23} & \val{55.83}{1.59} & \val{56.08}{14.96}& \val{39.50}{12.96}& \val{49.50}{5.90} & \val{35.68}{3.67} & \val{52.47}{5.75} & \val{38.29}{4.98} \\
CoL (Gemma4-31b)  & \val{80.66}{0.97} & \val{61.54}{0.58} & \val{65.21}{5.04} & \val{40.27}{2.62} & \val{73.04}{2.12} & \val{54.86}{1.79} & \val{81.29}{5.33} & \val{64.89}{4.23} \\
\midrule
\multicolumn{9}{l}{\textit{5-shot Setting}} \\
\midrule
HF (GPT-5.4-mini) & \val{64.78}{0.89} & \val{50.63}{0.96} & \val{13.01}{0.92} & \val{22.46}{2.48} & \val{27.94}{0.84} & \val{28.66}{1.01} & \val{64.24}{2.77} & \val{44.70}{2.09} \\
HF (Gemma4-31b)   & \val{66.55}{0.03} & \val{55.21}{0.07} & \val{12.08}{0.30} & \val{11.76}{0.46} & \val{29.89}{0.82} & \val{31.16}{1.03} & \val{54.83}{0.02} & \val{37.12}{0.17} \\
CoL (GPT-5.4-mini)& \val{80.66}{0.97} & \val{54.17}{1.08} & \val{52.90}{10.21}& \val{36.87}{10.05}& \val{59.40}{5.34} & \val{39.94}{5.97} & \val{64.16}{2.78} & \val{44.74}{2.15} \\
CoL (Gemma4-31b)  & \val{76.60}{9.50} & \val{60.10}{5.30} & \bval{77.99}{1.71}& \val{50.18}{3.14} & \val{78.81}{0.60} & \val{58.32}{0.92} & \val{80.71}{0.97} & \val{58.54}{1.63} \\
TIGA (GPT-5.4-mini) & \val{83.90}{4.46} & \val{62.97}{8.41} & \val{76.20}{8.35} & \bval{57.60}{8.25} & \val{71.90}{4.78} & \val{50.50}{5.27} & \val{69.00}{7.92} & \val{44.90}{1.80} \\
TIGA (Gemma4-31b)   & \bval{85.82}{2.83} & \bval{66.86}{5.81} & \val{75.60}{1.30} & \val{55.30}{2.80} & \bval{82.60}{2.20} & \bval{64.60}{1.70} & \bval{85.20}{0.76} & \bval{71.20}{1.56} \\
\bottomrule
\end{tabular}%
}
\end{table}
TIGA achieves the highest $F1_a$ or $F1_e$ in 7 of 8 reported settings. Its largest gains are concentrated in $R_e$ and $F1_e$: relative to the strongest baseline in each setting, TIGA improves $R_e$ by 8.1 percentage points on GProd, $R_e$ by 7.9 percentage points on DBLP, and $F1_e$ by 7.4 percentage points on DBLP. Improvements on ancestor-level metrics are more moderate but consistent, with increases of 3.8--5.3\% in $F1_a$ and $R_a$ on TExEval-2 and GProd.

TIGA does not dominate on every metric. On WordNet, DBLP, and GProd, it falls short of the baselines on some precision measures (e.g., $P_e$); the largest reported gap is 13.5 percentage points on DBLP $P_e$ relative to HF. The baselines' higher precision is accompanied by lower $R_e$, indicating a more conservative strategy that predicts fewer edges, but is more often correct when it does assign an entity to its corresponding hypernym parent. This precision--recall trade-off is reflected in HF's consistently lower $F1_e$ across the datasets. Compared with CoL, the stronger and more balanced prompting-based baseline, TIGA generally achieves higher recall and F1 across the ancestor- and edge-level measures, although the advantage is not uniform across every individual dataset--metric combination and reaches a deficit of at most 2 percentage points on DBLP's ancestor-level metrics.

Variance patterns are mixed rather than uniform: TIGA does not consistently achieve lower run-to-run variance than CoL. On GPT-5.4-mini, TIGA exhibits smaller standard deviations overall; however, on GProd it is less stable for $R_a$ and $F1_a$ (SD 8.5\% and 7.9\% versus 3.5\% and 2.8\% for CoL). Using the Gemma4-31b backbone, TIGA has larger standard deviations than CoL on 8 of the 18 reported metrics, with the largest disparity observed on GProd $P_a$ (4.2\% versus 0.8\%).

These results suggest that TIGA's multi-step construction is relatively stable on DBLP and TExEval-2 but introduces additional run-to-run variability on GProd. One plausible explanation is the larger candidate entity space in GProd, in which minor errors during early tool invocations may propagate and amplify over the longer interaction sequence. This interpretation remains tentative because the present experiments do not directly isolate the causal source of the observed increase in variability.

On WordNet, TIGA achieves the strongest $F1_a$ and $F1_e$ among the compared methods on both backbones. Backbone effects are method- and setting-dependent: moving from GPT-5.4-mini to Gemma4-31b increases $F1_a$ for HF (64.78\%\,$\to$\,66.55\%) and for TIGA (83.90\%\,$\to$\,85.82\%), but decreases $F1_a$ for CoL (80.66\%\,$\to$\,76.60\%). Overall, these results suggest that Gemma4-31b can improve WordNet performance; however, the gains are not uniform across prompting-based baselines. By contrast, TIGA remains the best-performing method under both evaluated backbones in the reported experimental setting. Nevertheless, the limited number of backbones does not permit a general claim of backbone independence.

Overall, the results provide substantial but not universal support for RQ1. TIGA achieves the strongest aggregate structural-quality results in most reported settings, particularly for recall and F1 scores at both the ancestor and edge levels. This relative advantage extends across different domains, taxonomy scales, and model backbones; however, its magnitude varies by benchmark and comes with lower precision and, in some settings, greater run-to-run variability.

\subsection{Exploratory Study of Frontier LLMs}
\label{app:flagship-robustness}
The experiments in Section~\ref{sec:baseline} use the cost-efficient backbones \texttt{GPT-5.4-mini} and \texttt{Gemma4-31b}, which provide competitive performance while keeping the computational and API budget managable for TIGA's multi-step construction process.
We further conduct a small indicative study with the current frontier models \texttt{GPT-5.6-luna} and \texttt{GPT-5.6-terra} to assess whether the observed trends extend to more capable LLMs.
We concentrate on five TExEval-2 sub-taxonomies in which TIGA shows the largest shortfall in $F1_a$ relative to CoL under the 5-shot condition with the \texttt{Gemma4-31b} backbone.

\begin{figure}[!htbp]
\centering
\includegraphics[width=\textwidth]{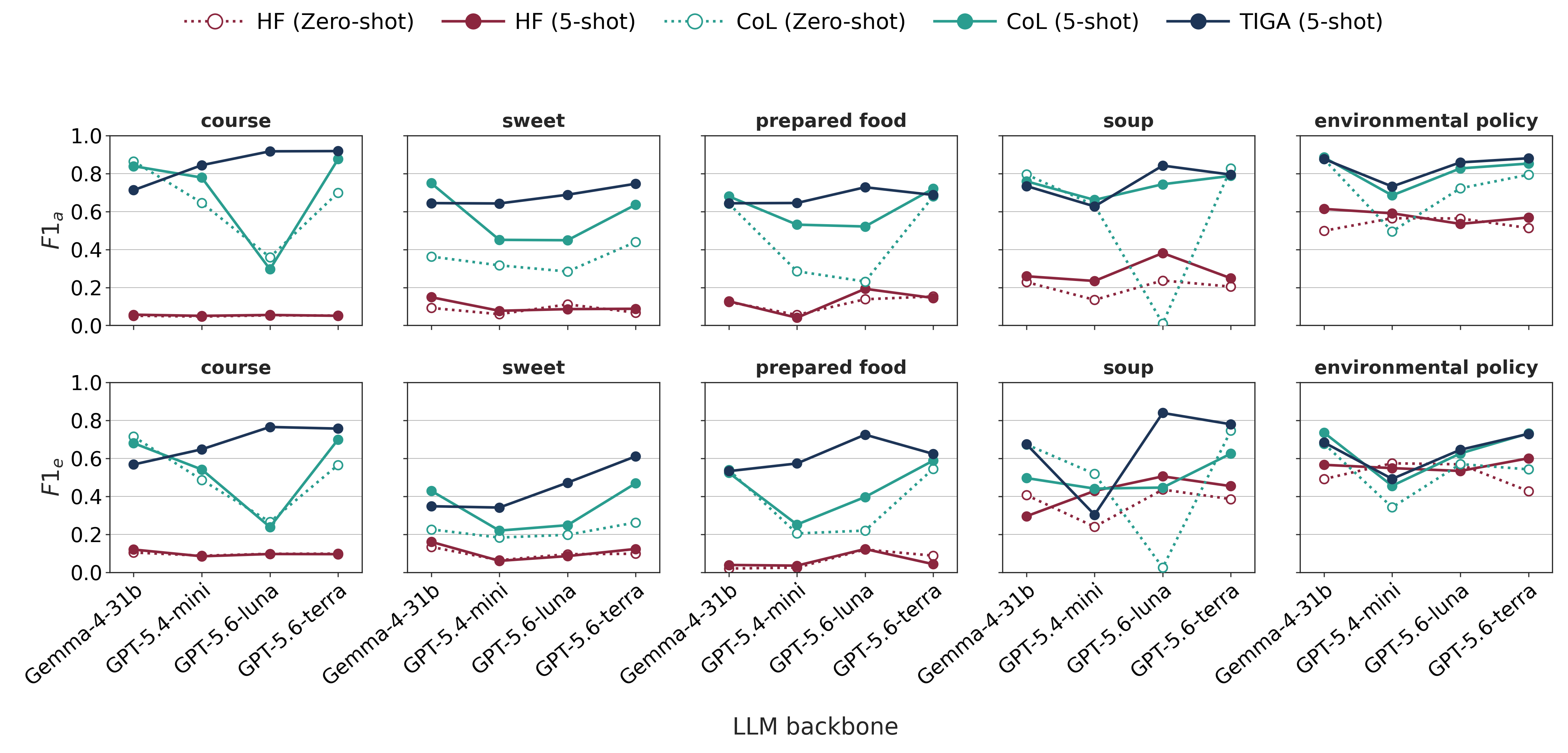}
\caption{Structural quality for $F1_a$ and $F1_e$ across four backbones, for 
five TExEval-2 sub-taxonomies ordered by largest deficit  
TIGA--CoL (5-shot) for Gemma4-31b.}
\label{fig:slope-chart-frontier}
\end{figure}

Fig.~\ref{fig:slope-chart-frontier} displays the five sub-taxonomies ordered from left (largest TIGA--CoL deficit) to right (smallest deficit). Along the x-axis, the backbones are ordered from the open-source \texttt{Gemma4-31b} to proprietary models, progressing from less capable to more capable.
Overall, TIGA exhibits more consistent results across different backbones, while CoL shows larger fluctuations on certain mid-range models (e.g., \textit{course} on \texttt{GPT-5.6-luna}, where CoL’s 5-shot $F1_a$ falls to 0.30 whereas TIGA remains at 0.92). In these cases, TIGA largely preserves its performance, and when moving to the next more capable backbone, it frequently matches or exceeds CoL. By contrast, HF remains relatively low on the selected sub-taxonomies and does not exhibit commensurate gains as backbone capability increases.

Moving from \texttt{GPT-5.6-luna} to \texttt{GPT-5.6-terra} results in only modest changes. For \textit{course}, TIGA stays essentially unchanged ($F1_a$ 0.917$\rightarrow$0.918), whereas for \textit{prepared food} and \textit{soup} it decreases on both $F1_a$ and $F1_e$. HF exhibits a similar saturation pattern on these sub-taxonomies, and CoL's luna--terra differences are small except where they reflect a prior drop. Overall, improvements from the largest model in this family are limited and not consistently positive across methods. Although this analysis is necessarily small and indicative, the results suggest partial convergence on stronger backbones, with reduced performance dispersion across interactive methods for several sub-taxonomies.

\subsection{Ablation Study}
\label{sec:ablation-study}

To evaluate the contribution of  architectural components, we conduct an ablation study in which we systematically disable components and run each configuration five times on the WordNet and TExEval-2 benchmarks.
Table~\ref{tab:ablation_components} reports the corresponding mean$_{\pm\text{std}}$ $F1_a$/$F1_e$ scores and absolute changes ($\Delta F1_a$, $\Delta F1_e$) relative to the baseline TIGA configuration.

\begin{table}[!htbp]
\centering
\caption{Component ablation study of TIGA on
WordNet and TExEval-2, using
Gemma-4-31b as backbone. Each row either removes or adds one component to method. Values are mean$_{\pm\text{std}}$ over 5 random
seeds; $\Delta F1_a$ and $\Delta F1_e$ report the
absolute percentage change relative to the baseline.}
\label{tab:ablation_components}
\small
\renewcommand{\arraystretch}{1.15}
\setlength{\tabcolsep}{8pt}
\begin{tabularx}{\columnwidth}{l c c c c c c c}
\toprule
Setting
& $\mathcal{P}_i$
& $M_i$
& $g_\phi$
& $F1_a$
& $\Delta F1_a$
& $F1_e$
& $\Delta F1_e$
\\
\midrule
\multicolumn{8}{l}{\textit{WordNet}}\\
-- Policy
& \xmark & \checkmark & \checkmark
& \val{82.26}{0.80} & $-3.56$
& \val{62.80}{1.00} & $-4.06$ \\
-- Memory
& \checkmark & \xmark & \checkmark
& \val{82.20}{0.50} & $-3.62$
& \val{61.30}{1.70} & $-5.56$ \\
-- Review
& \checkmark & \checkmark & \checkmark
& \val{86.90}{7.60} & $+1.18$
& \val{74.30}{1.70} & $+8.04$ \\
\midrule
\multicolumn{8}{l}{\textit{TExEval-2}}\\
-- Policy
& \xmark & \checkmark & \checkmark
& \val{79.92}{2.61} & $-2.68$
& \val{60.11}{3.32} & $-4.49$ \\
-- Memory
& \checkmark & \xmark & \checkmark
& \val{79.46}{3.73} & $-3.14$
& \val{58.80}{3.09} & $-5.80$ \\
-- Review
& \checkmark & \checkmark & \checkmark
& \val{81.20}{1.93} & $-1.40$
& \val{63.82}{2.35} & $-0.78$ \\
\bottomrule
\end{tabularx}
\end{table}

\begin{figure}[!htbp]
\centering
\includegraphics[width=0.9\columnwidth]{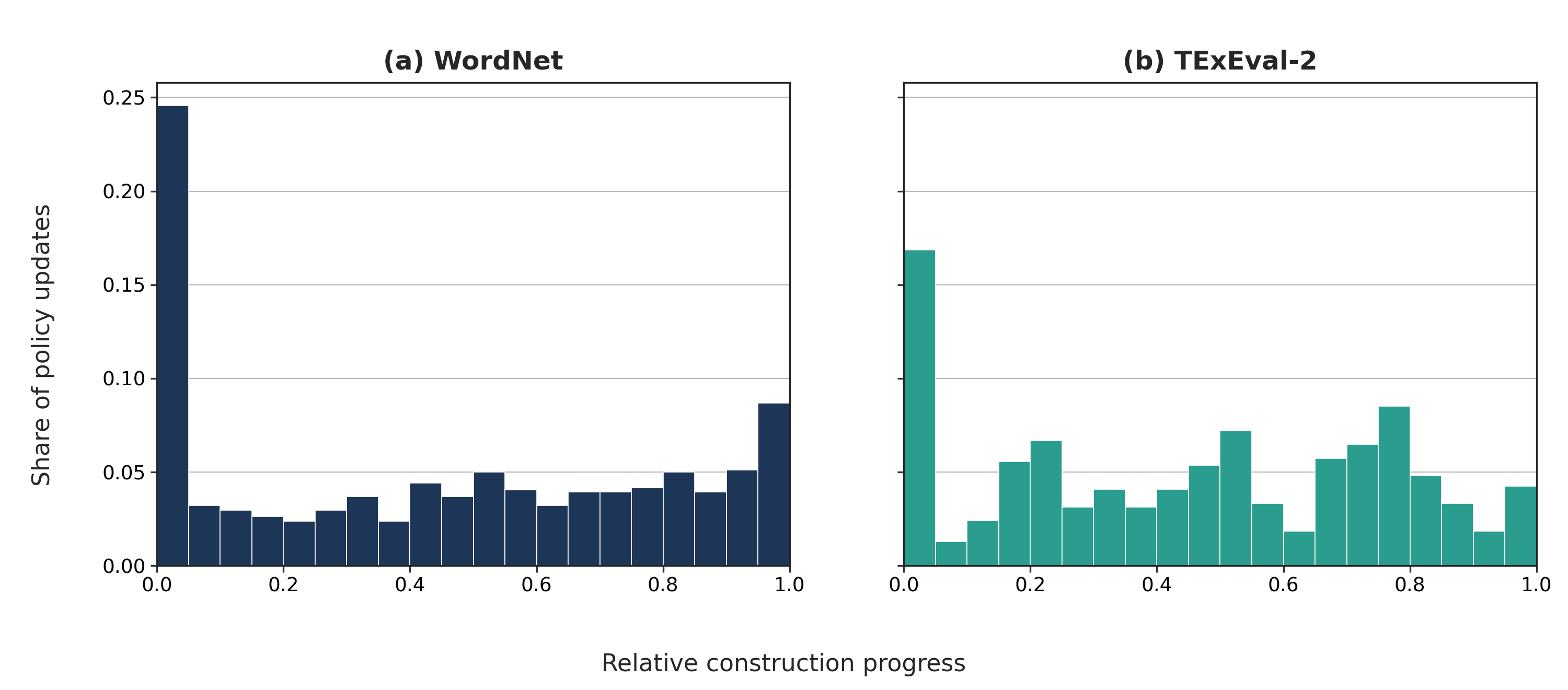}
\caption{Distribution of relative \texttt{update\_policy} call positions (0 = start, 1 = end of construction), averaged across runs by dataset. Both datasets show a bimodal pattern: an initial spike (policy establishment) and a terminal spike (consolidation), with a lower, approximately flat rate of mid-process correction.}
\label{fig:policy-timing-hist}
\end{figure}

Removing the policy module reduces $F1_a$ by 3.6\% on WordNet and 2.7\% on TExEval-2, and reduces $F1_{\text{e}}$ by 4.1\% and 4.5\%, respectively.
To characterize how TIGA uses the dynamic policy component, we analyze the relative positions of \texttt{update\_policy} calls along the construction trajectory for both datasets.
As illustrated in Fig.~\ref{fig:policy-timing-hist}, calls cluster within the first 5\% of construction (consistent with policy initialization), become infrequent and approximately uniform thereafter, with a modest uptick near the end that suggests late-stage consolidation. This pattern supports a two-stage use of the module: first, establishing organizing principles while the taxonomy is still sparse; then, selectively refining them once the overall structure is largely apparent.

TExEval-2 exhibits a clearer end-of-process spike and a higher mean call count than WordNet, which may reflect the additional planning and adjustment required by larger taxonomies (mean 3.79 calls/taxonomy on WordNet; 4.54 on TExEval-2; see Table ~\ref{tab:policy-timing-summary} for the full call-position distributions).
To illustrate how TIGA formalizes dynamic policy after initialization, Table~\ref{tab:policy-content-example} shows the policy from a TExEval-2 \textit{meat} sub-taxonomy run. It specifies a coarse-to-fine organization strategy, domain-specific groupings of meat cuts, and explicit exclusions.
Table~\ref{tab:policy-trace-example} traces all four \texttt{update\_policy} invocations from the same run. Updates progressively refine the organization, codify emerging rules, and add targeted exclusions. The trace further shows that the dynamic policy is invoked repeatedly as both a planning mechanism and a source of guidance throughout taxonomy construction, shaping subsequent decisions as the taxonomy evolves.

Eliminating the memory component yields a comparable degradation: $F1_a$ drops by 3.6\% on WordNet and 3.1\% on TExEval-2, while edge-level $F1_e$ drops by 5.6\% and 5.8\%, respectively. We attribute this effect to two potential benefits of the externalized memory: (i) it explicitly articulates the underlying rationale., thereby supporting decision consistency, and (ii) it enables the agent to condition on past induction errors, facilitating self-correction over time. Because the ablation removes the memory mechanisms jointly, these results do not isolate which of these two functions is responsible for the observed degradation.

Incorporating the reviewer increases ancestor-level $F1_a$ by 1.2\% and edge-level $F1_e$ by 8.0\% on WordNet, but slightly decreases both metrics on TExEval-2 (1.4\% and 0.8\%, respectively). To control computational costs, we capped the review procedure at three revision rounds. Under this cap, a non-negligible fraction of taxonomies in both benchmarks still fails the reviewer verdict after the third round and is returned without further modification. For WordNet's smaller, shallower sub-taxonomies, the three-round cap is rarely reached, and most issues are resolved within the available revision rounds. For TExEval-2, larger and more complex structures make three rounds less likely to suffice, which can leave repairs incomplete or introduce inconsistencies and thereby reduce performance relative to no review.

This finding indicates that, although the framework yields structural improvements in some settings, the resulting taxonomies remain imperfect. Aggregate gains in structural quality do not preclude individual erroneous taxonomies, and a reviewer verdict should not be interpreted as a guarantee of correctness. The capped review procedure can reveal and target some failures but cannot ensure their complete correction.

Taken together, the ablation results provide evidence that the policy and memory mechanisms contribute consistently to structural quality on the evaluated benchmarks. The benefit of the reviewer is dataset-dependent and appears to depend on whether the identified problems can be repaired within the bounded revision budget. These results provide a qualified answer to RQ2: all three components are useful in at least some evaluated settings, but their effects are not uniformly positive.

 \subsection{Computational Cost--Accuracy Trade-off}
 \label{sec:cost-quality}
\begin{figure}[!htbp]
\centering
\includegraphics[width=0.95\columnwidth,height=11cm,keepaspectratio]{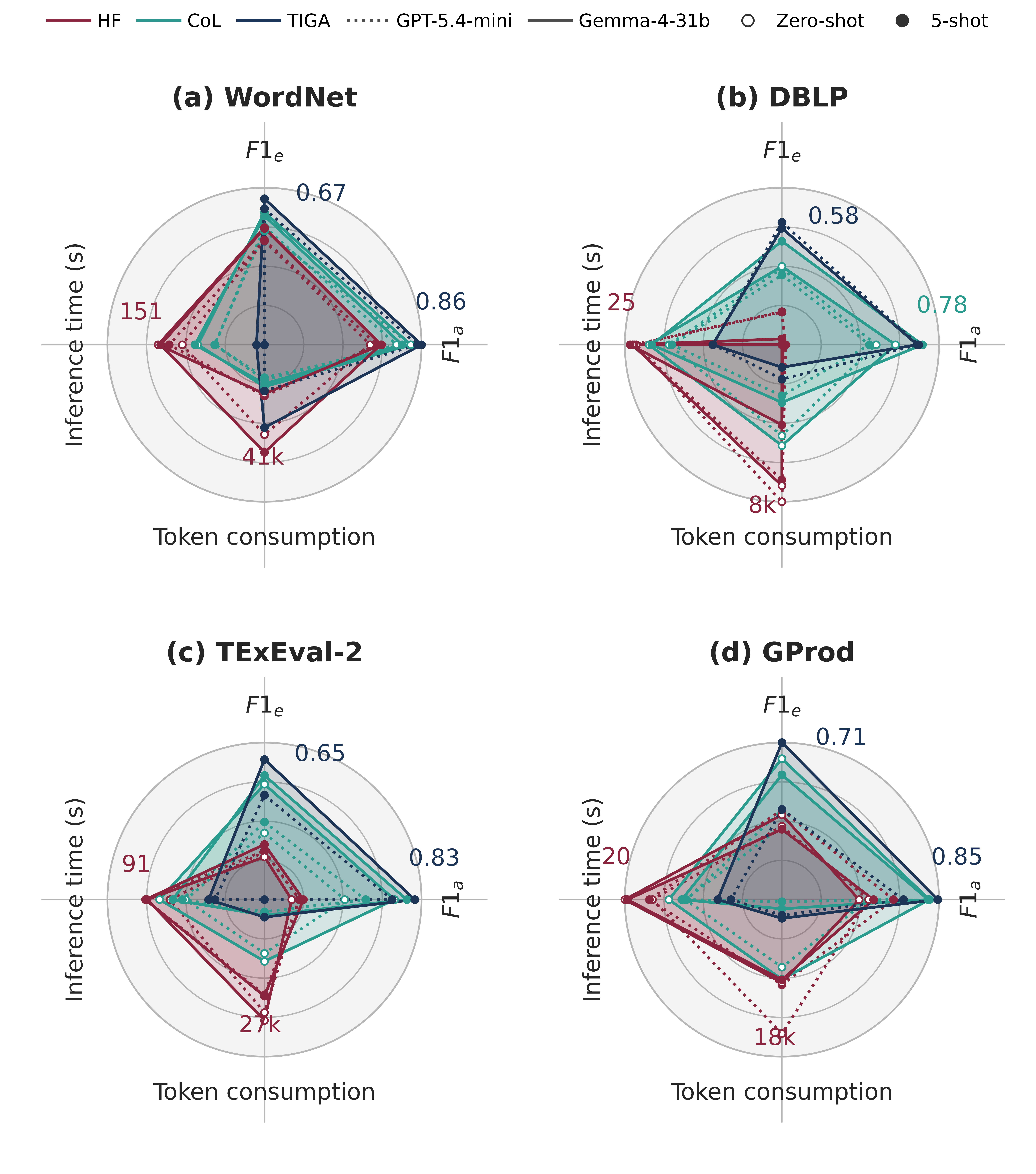}
\caption{Computational cost--quality trade-off per dataset visualized as a radar plot over four axes: $F1_e$, $F1_a$, total token consumption, and wall-clock inference time in seconds. Larger radial distance indicates better performance on all axes. For tokens and time, values are log-transformed, min--max normalized to $[0,1]$ across all configurations, and inverted ($1-\text{normalized}$) so that cheaper and faster runs map outward. Total tokens include both prompt and completion tokens. Color encodes method (HF, CoL, TIGA); line style encodes backbone model (solid = GPT-5.4-mini, dashed = Gemma4-31b); filled markers denote the 5-shot setting (or TIGA's default), and hollow markers denote the zero-shot setting.}
\label{fig:computational-tradeoff-radar}
\end{figure}

Fig.~\ref{fig:computational-tradeoff-radar} reports token consumption, API-call counts, and inference time in seconds, alongside $F1_a$ and edge-level $F1_e$, for each method/backbone/shot setting across the four datasets. In the radar plots, TIGA consistently lies in the highest-cost region. On TExEval-2 with GPT-5.4-mini, TIGA issues 1008.2 API calls to construct 18 sub-taxonomies, compared with 160.8 calls for CoL in the 5-shot setting (6.3$\times$ as many), and consumes 1.5$\times$ CoL's token budget for a comparable quality range ($F1_a$ 0.658 versus 0.594). A similar pattern appears on WordNet, where TIGA incurs higher API-call volume and longer inference time than both HF and CoL for each backbone.

This cost profile reflects how TIGA induces taxonomies: it uses fine-grained, tool-based edits on one entity per tool call and periodically exposes its plans through explicit policy updates. By contrast, HF's single-pass decoding and CoL's fixed layer-wise approach can modify multiple entities in a single call, while their intermediate reasoning remains implicit in the generated-output pipeline. Consequently, the number of TIGA calls increases with the number of entities and with the number of additional review or policy actions.

The findings indicate that the quality--cost trade-off is most favorable in settings where TIGA yields its largest structural-quality improvements. For WordNet and GProd, TIGA's gains in F1 are accompanied by higher computational overhead; whether they compensate for that overhead depends on the relative value assigned to taxonomy accuracy and computational resources. In contrast, on datasets where TIGA's structural-quality advantage is relatively small, the associated resource consumption is more difficult to justify. When taxonomy accuracy is the primary objective, the more expensive construction procedure may nevertheless be warranted to reduce the risk of error propagation in downstream applications, but this deployment consideration is application-dependent.

The choice of backbone model also substantially influences the cost profile. Within TIGA, Gemma4-31b is consistently more cost-effective than GPT-5.4-mini in the reported experiments while achieving comparable or superior quality. On TExEval-2, Gemma4-31b requires slightly fewer API calls (995.1 versus 1008.2) and attains a higher $F1_a$ (0.732 versus 0.658). On WordNet, it markedly reduces the number of API calls (2,656.3 versus 7,262.4) while further improving $F1_a$ (0.858 versus 0.839). This cost advantage appears specific to TIGA in the present experiment; for HF and CoL, the backbone choice primarily affects quality rather than call volume. Overall, the TIGA--Gemma4-31b configuration represents the most efficient reported setting under the combined cost--quality criterion used here.

The results provide a qualified answer to RQ3. TIGA achieves higher structural quality in many settings, but the improvement is associated with substantially higher token, call, and time consumption. Its practical suitability therefore depends on the required quality level, the size and complexity of the taxonomy, the available inference budget, and the cost of downstream errors.

\section{Conclusion}
\label{sec:conclusion}

This paper investigated whether an LLM-based agent can improve taxonomy induction by treating taxonomy construction as controlled knowledge engineering rather than unrestricted hierarchical text generation. We introduced TIGA, a tool-grounded framework in which an LLM invokes tool operations over an explicit taxonomy state, while a constrained execution layer applies and records the corresponding structural changes. The framework combines executable taxonomy-editing tools, memory, a runtime-adaptive organizational policy, and an optional global reviewer.

The experimental results provide qualified support for this design. Across four benchmark taxonomies and two LLM backbones, TIGA achieved the strongest structural-quality results in most evaluated configurations, with its clearest advantages observed in recall and F1. This pattern is consistent with the intended role of the architecture: the agent's explicit state, tool-mediated operations, and iterative feedback enable it to place a larger proportion of the candidate entities while maintaining competitive structural agreement with the gold taxonomies. The results do not, however, indicate uniform superiority. TIGA falls behind the baselines on some precision measures, exhibits additional variability on selected configurations, and is not equally effective across all domains and model backbones.

The ablation study further indicates that the dynamic policy and memory contribute consistently to structural quality on the evaluated benchmarks. The policy provides an explicit and revisable plan and control layer for organizational principles, while the memory preserves decisions and execution feedback that can support consistency and error-aware correction. The global reviewer has a more conditional effect: it improves some WordNet measures but slightly reduces performance on TExEval-2 under the bounded revision procedure. This result highlights that review is not automatically beneficial; its value depends on the complexity of the taxonomy, the quality of the detected revision targets, and whether the available revision budget is sufficient to repair them.

The cost analysis identifies an important limitation of the proposed approach. Fine-grained tool-mediated construction requires substantially more model interactions, tokens, and inference time than single-shot or layer-wise prompting. TIGA is therefore not a universal replacement for less expensive prompting-based methods. Its use is most defensible when the cost of incomplete, inconsistent, or difficult-to-audit taxonomies is high enough to justify additional inference resources. In this respect, TIGA should be understood as a quality--cost trade-off for knowledge construction rather than simply as an accuracy-improvement technique.

From a knowledge-based systems perspective, the main contribution is the explicit separation between semantic decision-making and the execution of knowledge-structure modifications. This separation makes intermediate decisions, rationales, tool outcomes, policy changes, and revision attempts inspectable and provides a basis for reproducible state transitions. It does not eliminate semantic errors or guarantee correct taxonomies, as the LLM remains responsible for selecting, executing, and justifying operations. Nevertheless, it offers a controllable architecture for integrating LLMs into knowledge-engineering workflows in which validity, traceability, and downstream reliability matter alongside predictive quality.

Several limitations define directions for future work. First, the evaluation uses a limited number of model backbones and benchmark taxonomies, and some datasets require pruning or conversion to fit the rooted-tree formulation. Broader evaluations should include additional domains, larger and deeper taxonomies, and native representations that preserve multiple-parent relations where appropriate. Second, the current memory design distinguishes an active context window from a retained interaction history, but the effects of different memory-retrieval and summarization strategies have not been isolated. Third, the reviewer uses a bounded revision budget and its decisions are not independently validated by human experts. Future work should investigate adaptive stopping criteria, stronger structural validators, human-in-the-loop review, and reviewer ensembles. 

Overall, TIGA demonstrates that runtime-adaptive policies and tool-grounded state transitions can make LLM-based taxonomy induction more explicit, inspectable, and structurally controllable. The results also show that these benefits come with non-trivial computational costs and remain dependent on the task and configuration. Future knowledge-based systems should therefore treat tool grounding, persistent state, and adaptive guidance not as guaranteed solutions to LLM reliability problems, but as design mechanisms whose value must be evaluated against the quality requirements and resource constraints of the target knowledge-engineering application.

\subsection*{Funding}
This research did not receive any specific grant from funding agencies in the public, commercial, or not-for-profit sectors. The open access option is supported by the DEAL contract between Elsevier and the University of Hamburg.

\subsection*{Declaration of generative AI and AI-assisted technologies in the software development and manuscript preparation process}
During the preparation of this work, the authors used OpenAI ChatGPT Codex and Cursor AI to generate and revise substantial portions of the research code from author-defined requirements and implementation logic, including Python code used to produce data visualizations from the experimental results using the jgraph Draw.io Plugin for Codex (version 1.1.0). AI assistance was also used to generate a repository for public release from the complete codebase and to improve the wording and readability of author-written manuscript text. The authors reviewed and edited the AI-assisted outputs and take full responsibility for the software, figures, and published paper.

\bibliographystyle{elsarticle-num-names}
\bibliography{bibliography/bibliography}

\clearpage
\FloatBarrier
\appendix
\section{Additional Results}
\FloatBarrier
\subsection{Dynamic Policy Updates}
\FloatBarrier

\label{app:policy-timing}

We log every \texttt{update\_policy} call across all completed 
5-shot TIGA runs (597 on WordNet, 82 on TExEval-2; 679 of 774 
logged runs completed) and record each call's relative position in 
the run (0 = start, 1 = end). Table~\ref{tab:policy-timing-summary} 
reports the mean and standard deviation of calls per taxonomy. Model identity is not recoverable 
from these logs, so this analysis pools across backbones; a 
backbone-stratified breakdown is left to future work.

\begin{center}
\captionof{table}{Mean \texttt{update\_policy} calls per taxonomy, by dataset}
\label{tab:policy-timing-summary}
\medskip
\begin{tabular}{lccc}
\toprule
Dataset & \#Runs & Mean calls/run & Std \\
\midrule
WordNet    & 597 & 3.79 & 1.90 \\
TExEval-2  &  82 & 4.54 & 2.73 \\
\bottomrule
\end{tabular}
\end{center}

\begin{center}
\captionof{table}{Excerpt of the agent-authored policy $\mathcal{P}_i$ for the TExEval-2 \textit{meat} sub-taxonomy, captured after the first \texttt{update\_policy} call. Rules are grouped by organizational category as authored by the agent. This snapshot pairs with Table~\ref{tab:policy-trace-example}, which traces how the policy is revised across the remaining construction process.}
\label{tab:policy-content-example}
\small
\begin{tabularx}{\columnwidth}{l X}
\toprule
\textbf{Category} & \textbf{Rule} \\
\midrule
Hierarchical structure & Taxonomy follows a logical progression from general to specific. \\
\midrule
Primary classification & Meat is first divided by processing state (fresh, frozen, tinned, deli) or by animal source (beef, pork). \\
\midrule
\multirow{2}{*}{Processing state vs.\ animal source}
& Terms describing meat state (e.g., ``frozen meat'') are top-level subcategories of \textit{meat}. \\
& Terms describing animal source (e.g., ``beef,'' ``pork'') are top-level subcategories of \textit{meat}. \\
\midrule
\multirow{4}{*}{Sausage classification}
& \textit{Sausage} is a distinct category. \\
& Specific sausage types (e.g., ``bratwurst,'' ``salami'') are subcategories of \textit{sausage}. \\
& \textit{Hot dogs} are a specific type of sausage. \\
& Specific hot dog styles (e.g., ``chicago-style hot dog'') are subcategories of \textit{hot dogs}. \\
\midrule
Specialized cuts & Specific cuts (e.g., ``pork chop,'' ``standing rib roast'') are placed under their respective animal source. \\
\midrule
By-products & Fats and suets are placed under the general \textit{meat} category, or under the specific animal where applicable. \\
\midrule
Exclusions & Processed items that are primarily \textit{sausage} are placed in the sausage subtree, even when made from beef or pork. \\
\bottomrule
\end{tabularx}
\end{center}

\label{app:policy-example}
Table~\ref{tab:policy-trace-example} shows the four 
\texttt{update\_policy} calls from a single representative run on 
TExEval-2 (the \textit{meat} sub-taxonomy), illustrating the 
progression from initial planning to late-stage correction 
discussed in Section~\ref{sec:ablation-study}.

\begin{center}
\captionof{table}{Policy updates obtained by TIGA execution on the TExEval-2 \textit{meat} sub-taxonomy. Each row corresponds to one \texttt{update\_policy} invocation, presented in chronological order.}
\label{tab:policy-trace-example}
\small
\begin{tabularx}{\columnwidth}{cX}
\toprule
\# & Update rationale (agent-generated) \\
\midrule
1 & Initial policy creation to guide the organization of a large batch of meat-related terms, ensuring a logical split between animal types, processing methods, and specific product categories like sausages. \\
2 & The initial batch contains a massive amount of sausages and specific meat cuts. A detailed policy is needed to distinguish between animal-source categories (beef/pork) and product-type categories (sausages), and to handle the specific sub-hierarchy of hot dogs. \\
3 & Formalizing the taxonomy principles based on the extensive organization of the meat, pork, beef, and sausage categories conducted in the previous run. \\
4 & Adding a specific principle to exclude processes from the taxonomy, as ``sausage making'' was identified as a process and not a product. \\
\bottomrule
\end{tabularx}
\end{center}

\subsection{Complete structural quality performance}
Table~\ref{tab:large_results} reports full precision, recall, and F1 
by dataset, method, and shot setting, under both ancestor-level 
($P_a$, $R_a$, $F1_a$) and edge-level ($P_e$, $R_e$, $F1_e$) protocols.

\begin{table}[htbp]
\centering
\tiny
\caption{Full precision, recall, and F1 by dataset, method, and 
shot setting, under both ancestor-level ($P_a$, $R_a$, $F1_a$) 
and edge-level ($P_e$, $R_e$, $F1_e$) protocols. Values are 
mean$_{\pm\text{std}}$ over 5 random seeds (in percent). Bold 
indicates the best result per column within each dataset.}
\label{tab:large_results}
\resizebox{0.97\columnwidth}{!}{%
\begin{tabular}{llcccccc}
\toprule
\textbf{Dataset} & \textbf{Method} & $P_a$ & $R_a$ & $F1_a$ & $P_e$ & $R_e$ & $F1_e$ \\
\midrule
\multirow{10}{*}{WordNet}
& \multicolumn{7}{l}{\textit{Zero-shot Setting}} \\
& HF (GPT-5.4-mini)  & \val{79.99}{1.08} & \val{54.66}{1.87} & \val{61.44}{1.76} & \val{60.00}{0.95} & \val{46.46}{1.32} & \val{50.32}{1.33} \\
& HF (Gemma4-31b)    & \val{84.01}{0.00} & \val{60.25}{0.06} & \val{66.83}{0.03} & \val{62.29}{0.16} & \val{50.98}{0.09} & \val{55.51}{0.11} \\
& CoL (GPT-5.4-mini) & \val{78.92}{2.04} & \val{72.16}{2.21} & \val{73.97}{2.23} & \val{57.42}{1.76} & \val{55.28}{1.61} & \val{55.83}{1.59} \\
& CoL (Gemma4-31b)   & \val{81.91}{1.20} & \val{81.07}{1.04} & \val{80.66}{0.97} & \val{61.85}{0.77} & \val{61.40}{0.58} & \val{61.54}{0.58} \\
\cmidrule(lr){2-8}
& \multicolumn{7}{l}{\textit{5-shot Setting}} \\
& HF (GPT-5.4-mini)  & \val{83.22}{0.52} & \val{58.33}{0.82} & \val{64.78}{0.89} & \val{61.06}{1.52} & \val{46.37}{1.00} & \val{50.63}{0.96} \\
& HF (Gemma4-31b)    & \val{83.70}{0.07} & \val{59.97}{0.04} & \val{66.55}{0.03} & \val{61.99}{0.13} & \val{50.72}{0.05} & \val{55.21}{0.07} \\
& CoL (GPT-5.4-mini) & \val{81.91}{0.61} & \val{81.07}{1.04} & \val{80.66}{0.97} & \val{54.57}{1.08} & \val{54.01}{1.07} & \val{54.17}{1.08} \\
& CoL (Gemma4-31b)   & \val{84.30}{0.50} & \val{73.70}{13.20}& \val{76.60}{9.50} & \val{64.10}{0.60} & \val{58.10}{7.80} & \val{60.10}{5.30} \\
& TIGA (GPT-5.4-mini)& \bval{88.22}{1.25}& \val{81.42}{0.38} & \val{83.90}{0.45} & \val{63.16}{1.18} & \val{62.88}{0.65} & \val{62.97}{0.84} \\
& TIGA (Gemma4-31b)  & \val{83.99}{2.78} & \bval{88.68}{2.67}& \bval{85.82}{2.83}& \bval{66.88}{5.80}& \bval{66.85}{5.82}& \bval{66.86}{5.81} \\
\midrule
\multirow{10}{*}{DBLP}
& \multicolumn{7}{l}{\textit{Zero-shot Setting}} \\
& HF (GPT-5.4-mini)  & \val{69.47}{4.01} & \val{7.42}{0.66}  & \val{13.28}{1.07} & \val{67.88}{4.57} & \val{13.54}{1.64} & \val{22.27}{2.36} \\
& HF (Gemma4-31b)    & \val{40.40}{0.00} & \val{6.74}{0.00}  & \val{11.49}{0.00} & \val{31.37}{0.00} & \val{5.61}{0.00}  & \val{9.45}{0.00} \\
& CoL (GPT-5.4-mini) & \val{59.27}{9.89} & \val{56.98}{16.69}& \val{56.08}{14.96}& \val{43.31}{9.88} & \val{37.73}{13.74}& \val{39.50}{12.96} \\
& CoL (Gemma4-31b)   & \val{55.94}{3.79} & \val{80.14}{7.45} & \val{65.21}{5.04} & \val{40.51}{2.58} & \val{40.05}{2.66} & \val{40.27}{2.62} \\
\cmidrule(lr){2-8}
& \multicolumn{7}{l}{\textit{5-shot Setting}} \\
& HF (GPT-5.4-mini)  & \val{67.86}{7.29} & \val{7.26}{0.50}  & \val{13.01}{0.92} & \bval{70.97}{7.29}& \val{13.50}{1.55} & \val{22.46}{2.48} \\
& HF (Gemma4-31b)    & \val{44.65}{1.97} & \val{7.05}{0.15}  & \val{12.08}{0.30} & \val{36.43}{0.95} & \val{7.10}{0.29}  & \val{11.76}{0.46} \\
& CoL (GPT-5.4-mini) & \bval{76.83}{6.29}& \val{48.22}{11.67}& \val{52.90}{10.21}& \val{47.84}{14.16}& \val{34.67}{9.42} & \val{36.87}{10.05} \\
& CoL (Gemma4-31b)   & \val{74.89}{2.09} & \bval{81.53}{1.83}& \bval{77.99}{1.71}& \val{50.57}{3.10} & \val{49.80}{3.18} & \val{50.18}{3.14} \\
& TIGA (GPT-5.4-mini)& \val{76.10}{8.99} & \val{76.90}{7.89} & \val{76.20}{8.35} & \val{57.50}{8.22} & \bval{57.70}{8.27}& \bval{57.60}{8.25} \\
& TIGA (Gemma4-31b)  & \val{74.80}{1.40} & \val{79.50}{2.00} & \val{75.60}{1.30} & \val{55.60}{2.60} & \val{54.90}{3.10} & \val{55.30}{2.80} \\
\midrule
\multirow{10}{*}{TExEval-2}
& \multicolumn{7}{l}{\textit{Zero-shot Setting}} \\
& HF (GPT-5.4-mini)  & \val{78.29}{6.46} & \val{20.87}{3.38} & \val{28.07}{3.77} & \val{62.81}{3.97} & \val{21.94}{3.07} & \val{28.31}{3.15} \\
& HF (Gemma4-31b)    & \bval{87.53}{0.02}& \val{16.33}{0.31} & \val{24.38}{0.35} & \val{65.86}{0.05} & \val{18.47}{0.24} & \val{26.25}{0.28} \\
& CoL (GPT-5.4-mini) & \val{72.16}{6.54} & \val{44.05}{5.47} & \val{49.50}{5.90} & \val{49.15}{5.58} & \val{32.18}{2.82} & \val{35.68}{3.67} \\
& CoL (Gemma4-31b)   & \val{77.36}{1.74} & \val{70.53}{2.54} & \val{73.04}{2.12} & \val{58.65}{1.95} & \val{52.21}{1.83} & \val{54.86}{1.79} \\
\cmidrule(lr){2-8}
& \multicolumn{7}{l}{\textit{5-shot Setting}} \\
& HF (GPT-5.4-mini)  & \val{85.67}{1.80} & \val{19.89}{0.66} & \val{27.94}{0.84} & \val{65.14}{4.02} & \val{21.38}{0.64} & \val{28.66}{1.01} \\
& HF (Gemma4-31b)    & \val{83.31}{1.12} & \val{20.82}{0.51} & \val{29.89}{0.82} & \val{61.46}{0.22} & \val{22.68}{0.87} & \val{31.16}{1.03} \\
& CoL (GPT-5.4-mini) & \val{79.67}{3.87} & \val{52.44}{4.60} & \val{59.40}{5.34} & \val{49.23}{5.00} & \val{36.58}{5.69} & \val{39.94}{5.97} \\
& CoL (Gemma4-31b)   & \val{83.47}{0.65} & \val{76.00}{1.56} & \val{78.81}{0.60} & \val{60.09}{1.24} & \val{56.92}{0.74} & \val{58.32}{0.92} \\
& TIGA (GPT-5.4-mini)& \val{80.30}{3.00} & \val{69.10}{5.84} & \val{71.90}{4.78} & \val{52.70}{4.31} & \val{50.20}{5.46} & \val{50.50}{5.27} \\
& TIGA (Gemma4-31b)  & \val{85.90}{1.80} & \bval{81.30}{2.40}& \bval{82.60}{2.20}& \bval{66.00}{1.70}& \bval{63.70}{1.80}& \bval{64.60}{1.70} \\
\midrule
\multirow{10}{*}{GProd}
& \multicolumn{7}{l}{\textit{Zero-shot Setting}} \\
& HF (GPT-5.4-mini)  & \val{71.70}{5.85} & \val{48.21}{5.71} & \val{52.53}{5.68} & \val{53.11}{3.85} & \val{34.31}{3.98} & \val{38.31}{4.95} \\
& HF (Gemma4-31b)    & \val{91.60}{0.00} & \val{35.58}{0.00} & \val{48.01}{0.00} & \bval{73.04}{0.00}& \val{32.98}{0.00} & \val{42.80}{0.00} \\
& CoL (GPT-5.4-mini) & \val{71.68}{5.78} & \val{48.24}{5.69} & \val{52.47}{5.75} & \val{53.14}{3.86} & \val{34.26}{4.02} & \val{38.29}{4.98} \\
& CoL (Gemma4-31b)   & \val{82.79}{3.98} & \val{81.51}{6.24} & \val{81.29}{5.33} & \val{68.17}{3.10} & \val{63.11}{4.87} & \val{64.89}{4.23} \\
\cmidrule(lr){2-8}
& \multicolumn{7}{l}{\textit{5-shot Setting}} \\
& HF (GPT-5.4-mini)  & \val{83.18}{1.30} & \val{58.32}{3.50} & \val{64.24}{2.77} & \val{54.52}{2.26} & \val{41.53}{2.80} & \val{44.70}{2.09} \\
& HF (Gemma4-31b)    & \bval{94.19}{0.77}& \val{41.65}{0.00} & \val{54.83}{0.02} & \val{54.76}{0.90} & \val{31.64}{0.13} & \val{37.12}{0.17} \\
& CoL (GPT-5.4-mini) & \val{83.15}{1.26} & \val{58.27}{3.46} & \val{64.16}{2.78} & \val{54.51}{2.19} & \val{41.48}{2.78} & \val{44.74}{2.15} \\
& CoL (Gemma4-31b)   & \val{87.35}{0.83} & \val{76.28}{1.52} & \val{80.71}{0.97} & \val{58.86}{1.57} & \val{58.22}{1.69} & \val{58.54}{1.63} \\
& TIGA (GPT-5.4-mini)& \val{74.70}{2.76} & \val{68.50}{8.53} & \val{69.00}{7.92} & \val{47.70}{1.96} & \val{45.40}{2.63} & \val{44.90}{1.80} \\
& TIGA (Gemma4-31b)  & \val{86.50}{4.20} & \bval{86.10}{0.98}& \bval{85.20}{0.76}& \val{71.10}{1.54} & \bval{71.20}{1.58}& \bval{71.20}{1.56} \\
\bottomrule
\end{tabular}%
}
\end{table}

\subsection{Computational Cost-Quality Trade-off}
Table~\ref{tab:consumption-tradeoff-radar-norm} summarizes per-combination metrics (edge-F1, ancestor-F1, token usage, API calls, and wall-clock time) for all (dataset, method, backbone, shot) configurations.

\begin{table}[htbp]
\centering
\tiny
\caption{Per-combination $F1_e$ and $F1_a$ (percent) and the corresponding
radar coordinates in $[0,1]$ (superscript n).
Further from the center is always better:
$F1_e^\mathrm{n}$/$F1_a^\mathrm{n}$ are min--max of the raw scores;
token and time coordinates are log-transformed, min--max normalized, then
reversed ($1-\mathrm{norm}$), so fewer tokens and lower runtime lie outward.
Ranges are global across all four datasets.
TIGA rows use the baseline configuration (5-shot equivalent).}
\label{tab:consumption-tradeoff-radar-norm}
\resizebox{\textwidth}{!}{%
\begin{tabular}{l l l l r r r r r r}
\toprule
Dataset & Method & Backbone & Shot & $F1_e$ & $F1_a$ & $F1_e^\mathrm{n}$ & $F1_a^\mathrm{n}$ & Tok.$^\mathrm{n}$ & Time$^\mathrm{n}$ \\
\midrule
\multirow{10}{*}{WordNet} & HF & GPT-5.4-mini & Zero-shot & 50.32 & 61.44 & 0.662 & 0.672 & 0.573 & 0.524 \\
 & HF & GPT-5.4-mini & 5-shot & 50.63 & 64.78 & 0.667 & 0.717 & 0.327 & 0.607 \\
 & HF & Gemma-4-31b & Zero-shot & 55.51 & 66.83 & 0.746 & 0.745 & 0.313 & 0.678 \\
 & HF & Gemma-4-31b & 5-shot & 55.21 & 66.55 & 0.741 & 0.741 & 0.685 & 0.661 \\
 & CoL & GPT-5.4-mini & Zero-shot & 55.83 & 73.97 & 0.751 & 0.841 & 0.226 & 0.315 \\
 & CoL & GPT-5.4-mini & 5-shot & 54.17 & 80.66 & 0.724 & 0.931 & 0.210 & 0.318 \\
 & CoL & Gemma-4-31b & Zero-shot & 61.54 & 80.66 & 0.844 & 0.931 & 0.268 & 0.424 \\
 & CoL & Gemma-4-31b & 5-shot & 60.10 & 76.60 & 0.820 & 0.876 & 0.249 & 0.443 \\
 & TIGA & GPT-5.4-mini & 5-shot & 62.97 & 83.90 & 0.867 & 0.974 & 0.294 & 0.000 \\
 & TIGA & Gemma-4-31b & 5-shot & 66.86 & 85.82 & 0.930 & 1.000 & 0.528 & 0.051 \\
\midrule
\multirow{10}{*}{DBLP} & HF & GPT-5.4-mini & Zero-shot & 22.27 & 13.28 & 0.208 & 0.024 & 1.000 & 0.928 \\
 & HF & GPT-5.4-mini & 5-shot & 22.46 & 13.01 & 0.211 & 0.020 & 0.861 & 0.941 \\
 & HF & Gemma-4-31b & Zero-shot & 9.45 & 11.49 & 0.000 & 0.000 & 0.897 & 0.953 \\
 & HF & Gemma-4-31b & 5-shot & 11.76 & 12.08 & 0.037 & 0.008 & 0.511 & 0.967 \\
 & CoL & GPT-5.4-mini & Zero-shot & 39.50 & 56.08 & 0.487 & 0.600 & 0.580 & 0.719 \\
 & CoL & GPT-5.4-mini & 5-shot & 36.87 & 52.90 & 0.444 & 0.557 & 0.325 & 0.701 \\
 & CoL & Gemma-4-31b & Zero-shot & 40.27 & 65.21 & 0.499 & 0.723 & 0.642 & 0.846 \\
 & CoL & Gemma-4-31b & 5-shot & 50.18 & 77.99 & 0.660 & 0.895 & 0.368 & 0.828 \\
 & TIGA & GPT-5.4-mini & 5-shot & 57.60 & 76.20 & 0.780 & 0.871 & 0.219 & 0.439 \\
 & TIGA & Gemma-4-31b & 5-shot & 55.30 & 75.60 & 0.743 & 0.863 & 0.144 & 0.441 \\
\midrule
\multirow{10}{*}{TExEval-2} & HF & GPT-5.4-mini & Zero-shot & 28.31 & 28.07 & 0.305 & 0.223 & 0.719 & 0.601 \\
 & HF & GPT-5.4-mini & 5-shot & 28.66 & 27.94 & 0.311 & 0.221 & 0.605 & 0.756 \\
 & HF & Gemma-4-31b & Zero-shot & 26.25 & 24.38 & 0.272 & 0.173 & 0.769 & 0.748 \\
 & HF & Gemma-4-31b & 5-shot & 31.16 & 29.89 & 0.352 & 0.248 & 0.615 & 0.758 \\
 & CoL & GPT-5.4-mini & Zero-shot & 35.68 & 49.50 & 0.425 & 0.511 & 0.342 & 0.502 \\
 & CoL & GPT-5.4-mini & 5-shot & 39.94 & 59.40 & 0.494 & 0.645 & 0.077 & 0.523 \\
 & CoL & Gemma-4-31b & Zero-shot & 54.86 & 73.04 & 0.735 & 0.828 & 0.393 & 0.668 \\
 & CoL & Gemma-4-31b & 5-shot & 58.32 & 78.81 & 0.791 & 0.906 & 0.100 & 0.584 \\
 & TIGA & GPT-5.4-mini & 5-shot & 50.50 & 71.90 & 0.665 & 0.813 & 0.000 & 0.315 \\
 & TIGA & Gemma-4-31b & 5-shot & 64.60 & 82.60 & 0.893 & 0.957 & 0.111 & 0.356 \\
\midrule
\multirow{10}{*}{GProd} & HF & GPT-5.4-mini & Zero-shot & 38.31 & 52.53 & 0.467 & 0.552 & 0.853 & 0.824 \\
 & HF & GPT-5.4-mini & 5-shot & 44.70 & 64.24 & 0.571 & 0.710 & 0.543 & 0.844 \\
 & HF & Gemma-4-31b & Zero-shot & 42.80 & 48.01 & 0.540 & 0.491 & 0.529 & 1.000 \\
 & HF & Gemma-4-31b & 5-shot & 37.12 & 54.83 & 0.448 & 0.583 & 0.510 & 0.985 \\
 & CoL & GPT-5.4-mini & Zero-shot & 38.29 & 52.47 & 0.467 & 0.551 & 0.430 & 0.615 \\
 & CoL & GPT-5.4-mini & 5-shot & 44.74 & 64.16 & 0.571 & 0.709 & 0.013 & 0.608 \\
 & CoL & Gemma-4-31b & Zero-shot & 64.89 & 81.29 & 0.898 & 0.939 & 0.507 & 0.720 \\
 & CoL & Gemma-4-31b & 5-shot & 58.54 & 80.71 & 0.795 & 0.931 & 0.059 & 0.638 \\
 & TIGA & GPT-5.4-mini & 5-shot & 44.90 & 69.00 & 0.574 & 0.774 & 0.099$^{\dagger}$ & 0.324$^{\dagger}$ \\
 & TIGA & Gemma-4-31b & 5-shot & 71.20 & 85.20 & 1.000 & 0.992 & 0.119 & 0.409 \\
\bottomrule
\end{tabular}%
}
\vspace{0.4em}
{\tiny $^{\dagger}$ Token/time coordinates use imputed infra (F1 measured).}
\end{table}

\section{Tool Functions}
Table~\ref{tab:tools} specifies the tool interface exposed to the agent via the LLM tool-calling interface. All tools require a \texttt{rationale} argument, whose value is appended to the memory $M_i$ as part of the tuple $(a_i, \rho_i, o_i)$ upon execution. State-mutating tools (\textit{modify}) produce a successor state $S_{i+1} \neq S_i$ on success; the read-only tool (\textit{inspect}) leaves $S_i$ unchanged. On failure ($o_i = \textsc{err}$), no structural modification is committed to $T_i$ or $V_i$; the error is recorded in $M_{i+1}$.

\begin{center}
\begin{minipage}{\linewidth}
\centering
\renewcommand{\arraystretch}{1.45}
\setlength{\tabcolsep}{6pt}
\captionof{table}{Tool interface specifications exposed to the agent via the LLM tool-calling interface. All tools require a \texttt{rationale} argument, whose value is appended to the memory $M_i$ as part of the tuple $(a_i, \rho_i, o_i)$ upon execution. State-mutating tools (\textit{modify}) produce a successor state $S_{i+1} \neq S_i$ on success; the read-only tool (\textit{inspect}) leaves $S_i$ unchanged. On failure ($o_i = \textsc{err}$), no structural modification is committed to $T_i$ or $V_i$; the error is recorded in $M_{i+1}$.}
\label{tab:tools}
\medskip
\small
\begin{tabular}{%
  l                  
  l                  
  l                  
  l                  
  l                  
}
\toprule
\textbf{Tool}
  & \textbf{Class}
  & \textbf{Arguments}
  & \textbf{Returns}
  & \textbf{Effect on $S_i$} \\
\midrule
\texttt{add\_entity}
  & \textit{modify}
  & \begin{tabular}[t]{@{}l@{}}
      \texttt{parent\_id : int} \\
      \texttt{entity\_name : str} \\
      \texttt{rationale : str}
    \end{tabular}
  & \begin{tabular}[t]{@{}l@{}}
      $o_i \in \{\textsc{ok},\textsc{err}\}$ \\
      \textsc{ok}: new node id
    \end{tabular}
  & \begin{tabular}[t]{@{}l@{}}
      $T_{i+1}$: node appended \\
      $V_{i+1}$: entity resolved \\
      $M_{i+1}$: tuple appended
    \end{tabular} \\
\midrule
\texttt{remove\_entity}
  & \textit{modify}
  & \begin{tabular}[t]{@{}l@{}}
      \texttt{entity\_id : int} \\
      \texttt{entity\_name : str} \\
      \texttt{rationale : str}
    \end{tabular}
  & \begin{tabular}[t]{@{}l@{}}
      $o_i \in \{\textsc{ok},\textsc{err}\}$ \\
      \textsc{err}: node not found \\
      \phantom{\textsc{err}:} or root removal
    \end{tabular}
  & \begin{tabular}[t]{@{}l@{}}
      $T_{i+1}$: node + descendants \\
      \phantom{$T_{i+1}$:} removed \\
      $V_{i+1}$: entity resolved \\
      $M_{i+1}$: tuple appended
    \end{tabular} \\
\midrule
\texttt{view\_subtree}
  & \textit{inspect}
  & \begin{tabular}[t]{@{}l@{}}
      \texttt{entity\_id : int}$^{\ddagger}$ \\
      \texttt{rationale : str}
    \end{tabular}
  & \begin{tabular}[t]{@{}l@{}}
      Descendant subgraph \\
      of $T_i$ in markdown
    \end{tabular}
  & \begin{tabular}[t]{@{}l@{}}
      $S_{i+1} = S_i$ \\
      $M_{i+1}$: tuple appended
    \end{tabular} \\
\midrule
\texttt{update\_policy}
  & \textit{modify}
  & \begin{tabular}[t]{@{}l@{}}
      \texttt{new\_policy\_content : str} \\
      \texttt{rationale : str}
    \end{tabular}
  & \begin{tabular}[t]{@{}l@{}}
      $o_i \in \{\textsc{ok},\textsc{err}\}$ \\
      \textsc{ok}: policy persisted
    \end{tabular}
  & \begin{tabular}[t]{@{}l@{}}
      $\mathcal{P}_{i+1}$: replaced \\
      $M_{i+1}$: tuple appended
    \end{tabular} \\
\bottomrule
\end{tabular}

\smallskip
{\raggedright\footnotesize
$^{\ddagger}$~Setting \texttt{entity\_id = 0} returns the full taxonomy rooted at $T_i$.\par
}
\end{minipage}
\end{center}

\end{document}